\documentclass[epsfig,floats,pre,twocolumn,showpacs,preprintnumbers,floatfix]{revtex4-1}
\usepackage{graphicx}
\usepackage[latin2]{inputenc}
\usepackage{amsmath}
\usepackage{amssymb}
\usepackage{bm}
\usepackage{color}

\newcommand{\Fvi}{F\!\!_{_{\text{I}}}{(v)}}
\newcommand{\Fvf}{F\!\!_{_{\text{I\!I}}}{(v)}}
\newcommand{\vj}{v_{_{\text{j}}}}
\newcommand{\vjm}{\langle v\rangle_{_{\text{j}}}}
\newcommand{\vjsm}{\langle v^2\rangle_{_{\text{j}}}}

\newcommand{\vjmp}{\langle v\rangle_{_{\text{j}'}}}

\newcommand{\ai}{a_{_{\text{I}}}}
\newcommand{\af}{a_{_{\text{I\!I}}}}
\newcommand{\aFif}{a_{_{\text{I}\rightarrow\text{I\!I}}}}
\newcommand{\aFfi}{a_{_{\text{I\!I}\rightarrow\text{I}}}}
\newcommand{\Ri}{R_{_{\text{I}}}\!(\phi)}
\newcommand{\Rf}{R_{_{\text{I\!I}}}\!(\phi)}
\newcommand{\aif}{R_{_{\text{I}\rightarrow\text{I\!I}}}\!(\phi)}
\newcommand{\afi}{R_{_{\text{I\!I}\rightarrow\text{I}}}\!(\phi)}
\newcommand{\aj}{a_{\text{j}}}
\newcommand{\ajp}{a_{_{\text{j}'}}}

\newcommand{\fif}{f\!_{_{\text{I}\rightarrow\text{I\!I}}}}
\newcommand{\ffi}{f\!_{_{\text{I\!I}\rightarrow\text{I}}}}
\newcommand{\fjjp}{f\!\!_{_{_{\text{j}{\rightarrow}\text{j}'}}}}
\newcommand{\fjpj}{f\!\!_{_{_{\text{j}'{\rightarrow}\text{j}}}}}

\newcommand{\Pti}{P\!\!_t^{\;\,\text{I}}(x,y|\gamma)}
\newcommand{\Ptf}{P\!\!_t^{\;\,\text{I\!I}}(x,y|\gamma)}

\newcommand{\Ptm}{\left(\!\!\begin{array}{ccccccccc}
\Pti \\ & \\ \Ptf\end{array}\!\!\!\!\!\!\right)}
\newcommand{\Ptpi}{P\!\!_{t{-}\Delta t}^{\;\,\text{I}}(x',y'|\beta)}
\newcommand{\Ptpf}{P\!\!_{t{-}\Delta t}^{\;\,\text{I\!I}}(x',y'|\beta)}
\newcommand{\Ptpm}{\left(\!\!\begin{array}{ccccccccc}
\Ptpi \\ & \\ \Ptpf\end{array}\!\!\!\!\!\!\right)}

\newcommand{\xp}{x{-}v \Delta t\cos\gamma}
\newcommand{\yp}{y{-}v \Delta t\sin\gamma}

\newcommand{\dphi}{\int_{-{\pi}}^{{\pi}}\!\text{d}\phi}
\newcommand{\Mtran}{\displaystyle\int \!\!\! dv \!\!\!
\displaystyle\int_{\!{-}\pi}^{\pi} \!\!\!\!\!\! d\beta 
\!\!\displaystyle\left[ \!\!\!\begin{array}{ccccccccc}
(1{-}\fif\!) \Fvi R_{_{\text{I}}}\!(\gamma{-}\beta) & 
\ffi \Fvi R_{_{\text{I\!I}\rightarrow\text{I}}}\!(\gamma{-}\beta) \\
& \\
\!\!\!\!\fif \Fvf R_{_{\text{I}\rightarrow\text{I\!I}}}\!(\gamma{-}\beta) & 
\!\!\!\!(1{-}\ffi\!) \Fvf R_{_{\text{I\!I}}}\!(\gamma{-}\beta)
\end{array} \!\!\!\right]}

\newcommand{\ts}{t_\text{s}}
\newcommand{\tcp}{t_{\text{c}_{+}}}
\newcommand{\tcm}{t_{\text{c}_{-}}}

\newcommand{\ftone}{R_{_{_{\text{1}}}}\!(\gamma{-}\beta)}
\newcommand{\ftn}{R_{_{_{n}}}\!(\gamma{-}\beta)}
\newcommand{\qonej}{f\!_{_{\text{1}\rightarrow\text{j}}}}
\newcommand{\sqonej}{\sum\limits_{j{\neq}1}\!\!\qonej}
\newcommand{\qtwoone}{f\!_{_{\text{2}\rightarrow\text{1}}}}

\newcommand{\qnone}{f\!_{_{n\rightarrow\text{1}}}}
\newcommand{\qonen}{f\!_{_{\text{1}\rightarrow n}}}
\newcommand{\qtwon}{f\!_{_{\text{2}\rightarrow n}}}
\newcommand{\qnj}{f\!_{_{n\rightarrow\text{j}}}}
\newcommand{\sqnj}{\sum\limits_{j{\neq}n}\!\!\qnj}
\newcommand{\Snewone}{P_{\!_{t}}^{\,\text{1}}(x,y|\gamma)}
\newcommand{\Snewn}{P_{\!_{t}}^{\,n}(x,y|\gamma)}
\newcommand{\Soldone}{P_{t{-}\Delta t}^{\,\text{1}}(x',y'|\beta)}
\newcommand{\Soldn}{P_{t{-}\Delta t}^{\,n}(x',y'|\beta)}
\newcommand{\SnewVec}{\left(\!\!\begin{array}{ccccccccc}
\Snewone \\ \vdots \\ \Snewn \end{array}\!\!\right)}
\newcommand{\SoldVec}{\left(\!\!\begin{array}{ccccccccc}
\Soldone \\ \vdots \\ \Soldn \end{array}\!\!\right)}

\newcommand{\MtranNN}{\!\displaystyle\int \!\!\! dv\!\!\displaystyle\int_{\!{-}\pi}^{\pi} \!\!\!\!\!\! 
d\beta \!\!\displaystyle\left[\!\!
\begin{array}{cccc}
(1{-}\!\!\!\sqonej\!)\ftone F\!\!_{_{_{\text{1}}}}\!(v)\,&\!\!\qtwoone R_{_{_{\text{2}\rightarrow\text{1}}}}\!(\gamma{-}\beta) F\!\!_{_{_{\text{1}}}}\!(v)\,&\!\!\cdots\,&\!\!\qnone R_{_{_{n\rightarrow\text{1}}}}\!(\gamma{-}\beta) F\!\!_{_{_{\text{1}}}}\!(v)\\
\vdots\,&\!\!\ddots\,& \,&\!\!\vdots\\
\!\!\!\!\!\!\!\!\qonen R_{_{_{\text{1}\rightarrow n}}}\!(\gamma{-}\beta) F\!\!_{_{_{n}}}\!(v)\,&\!\!\!\qtwon R_{_{_{\text{2}\rightarrow n}}}\!(\gamma{-}\beta) F\!\!_{_{_{n}}}\!(v)\,&\!\!\!\!\!\cdots\,&\!\!\!\!\!(1{-}\!\!\!\!\sqnj\!)\ftn F\!\!_{_{_{n}}}\!(v)
\end{array}
\!\!\!\right]\!\!\!}

\newcommand{\vf}{\langle v\rangle_{_{\text{I\!I}}}}
\newcommand{\vi}{\langle v\rangle_{_{\text{I}}}}
\newcommand{\Vdoi}{\langle v^2\rangle_{_{\text{I}}}}
\newcommand{\Vdof}{\langle v^2\rangle_{_{\text{I\!I}}}}

\begin{document}
\title{Kinematics of Persistent Random Walkers with Two Distinct Modes of Motion}
\author{M.\ Reza Shaebani}
\email{shaebani@lusi.uni-sb.de}
\author{Heiko Rieger}
\author{Zeinab Sadjadi}
\email{sadjadi@lusi.uni-sb.de}
\affiliation{Department of Theoretical Physics $\&$ Center for Biophysics, 
Saarland University, 66123 Saarbr\"ucken, Germany}

\begin{abstract}
We study the stochastic motion of active particles that undergo spontaneous 
transitions between two distinct modes of motion. Each mode is characterized 
by a velocity distribution and an arbitrary (anti-)persistence. We present an 
analytical formalism to provide a quantitative link between these two microscopic 
statistical properties of the trajectory and macroscopically observable transport 
quantities of interest. For exponentially distributed residence times in each 
state, we derive analytical expressions for the initial anomalous exponent, 
the characteristic crossover time to the asymptotic diffusive dynamics, and 
the long-term diffusion constant. We also obtain an exact expression for 
the time evolution of the mean square displacement over all time scales 
and provide a recipe to obtain higher displacement moments. Our approach 
enables us to disentangle the combined effects of velocity, persistence, 
and switching probabilities between the two states on the kinematics of 
particles in a wide range of stochastic active/passive processes and to 
optimize the transport quantities of interest with respect to any of the 
particle dynamics properties.
\end{abstract}



\maketitle

\section{Introduction}
Transport processes and active motion in nature often consist of more than one 
motility state. Examples in biological systems include swimming of bacteria 
\cite{Berg04}, migration of dendritic cells \cite{Chabaud15}, chemical signal 
transport in neuronal dendrites \cite{Jose18}, searching for specific target 
sites by DNA-binding proteins \cite{Bauer12,Meroz09}, growth of biopolymers 
\cite{Dogterom93,Shaebani16}, and motion of molecular motors along cytoskeleton 
\cite{Klumpp05}. While two-state transport processes are ubiquitous in nature, 
stochastic processes with multiple states have also been observed in natural 
systems, such as the three-state motion of {\it E.\,coli} near surfaces 
\cite{PerezIpina19}.

The dynamics of active particles is often described by the interplay between 
propulsion and stochastic forces. However, the origin of the exerted forces 
may be unknown in general. Moreover, the particle dynamics does not necessarily 
always originate from external fields--- e.g.\ in robotics the occurrence of 
reorientation events is controlled by the internal robot dynamics \cite{Nava18,
Nava21} and the motion should be described by connecting the space of internal 
states of the robot to the physical space in which it moves. Instead of 
describing the particle dynamics based on the exerted forces, one can alternatively 
obtain the macroscopically observable transport quantities of interest from 
the microscopic statistical properties of the trajectory, such as the velocity 
and turning-angle distributions of the particle and the transition probabilities 
between the possible states of motion. The solvability of such models however depends 
on the mathematical form of the statistics of the particle trajectories (e.g.\ its 
turning-angle distribution), which usually restricts the analysis to specific regimes 
of motion such as the asymptotic long-term dynamics. Nevertheless, the intermediate- 
and short-time regimes of motion are often of particular interest and the observation 
time of (biological) experiments is typically not long enough to realize the long-term 
regime. Moreover, the information extracted from trajectories--- such as the 
turning-angle distribution--- may have a complex form in general. Thus, a general 
formalism to derive the time evolution of the transport quantities over all 
timescales for arbitrary forms of the particle trajectory statistics is required 
which is technically challenging.

To model active dynamics with multiple states, simple combinations of 
stochastic processes--- such as a ballistic flight and a diffusion--- 
have been widely employed to capture some of the specific features of these 
systems \cite{Bressloff13,Hafner16,Taktikos13,Pinkoviezky13,Shaebani18,
Watari10,Theves13,Shaebani19}. A pertinent example is the bacterial dynamics, 
often modeled as ballistic run phases interrupted by periods of diffusion 
or random reorientation events, the so-called {\it run-and-tumble} dynamics 
\cite{Angelani09,Elgeti15,Thiel12,Angelani13,VillaTorrealba20}. The run 
trajectories are however curved (even spiral trajectories may form near 
surfaces \cite{PerezIpina19}) and the run-phase persistence, duration, 
and velocity vary with structural properties of bacteria or in response to 
environmental changes \cite{Berg04,Patteson15,Najafi18}. Also, tumbling 
is not a diffusion but rather an active phase with a reduced persistence; 
the flagellar bundles are only partially disrupted and there remains a weak 
swimming power to proceed forward \cite{Najafi18,Turner16}. Therefore, a 
simplified ballistic-diffusive model for the bacterial dynamics is inadequate. 
A full description of such multistate stochastic processes requires a more 
complete formalism to consider underlying correlations and memory effects 
and combine multiple states of motion with arbitrary persistencies and velocities, 
and general transitions between the states. 

Here we present a theoretical approach to combine two distinct stochastic 
processes with arbitrary persistencies and velocity distributions. The formalism 
can be extended to processes with multiple states in general. To be analytically 
tractable, we ignore possible underlying correlations and memory effects and 
consider spontaneous transitions between the two states, leading to exponential 
residence times in each state. The formalism is capable of handling correlations 
in general, though one should then resort to numerical methods to extract the 
transport quantities. We adopt a discrete-time process to be directly applicable 
to the experimental data--- which usually consist in particle positions recorded 
at equidistant times. We derive an exact expression for the time evolution of 
the mean square displacement (MSD) over all timescales (with the lower time 
resolution being limited by the frequency of the particle position recording, 
i.e.\ the camera frame rate, in experiments). We also provide the recipe to 
calculate higher displacement moments. The formalism presented here enables 
us to link macroscopically observable transport properties, such as the asymptotic 
diffusion coefficient $D\!_{_\infty}$, to two microscopic statistical properties 
of the trajectory: the velocity distribution and persistence.

Functioning in an optimal way is ubiquitously observed in biological systems. A 
pertinent example is the optimization of the escape or search times \cite{PerezIpina19,
Bauer12,Schuss07,Bartumeus08}. Minimizing search times often corresponds to maximizing 
the asymptotic diffusion coefficient $D\!_{_\infty}$, as they are conversely 
related to each other \cite{Condamin05,Redner01}. We show how $D\!_{_\infty}$ depends 
on several key factors, including the velocity and persistence of each state 
and the switching statistics between the two states. In practice, a biological agent 
may be able to vary only a few of these parameters. For instance, bacteria can adapt 
their run persistence or run-to-tumble switching frequency to enhance their diffusivity 
\cite{Berg04}. By obtaining the derivative of $D\!_{_\infty}$ with respect to any 
influential parameter and maximizing it, we can verify whether an optimal diffusion 
coefficient can be achieved by varying that specific parameter and how much the diffusivity 
can be enhanced.

\section{Model}
\label{Sec:Model}
We consider a stochastic active process with two different modes $\text{I}$ and 
$\text{I\!I}$ of motion. Each mode is characterized by the statistics of its 
velocity and persistence, as described below. Whereas a two-state 
process is chosen as the most frequent multi-state process in natural systems, 
the formalism can be generalized to processes with multiple states. We also 
note that a 2D active motion is studied here for brevity but nonetheless 
extension to 3D is straightforward (see e.g.\ \cite{Sadjadi15} for 3D 
treatment of a single-state active process). Using a discrete-time approach, 
we describe the motion by a discrete set of particle positions recorded 
after successive time intervals of size $\Delta t$. By setting the timestep 
$\Delta t$ to the inverse frame rate of the camera in experiments, our 
formalism and results can be directly applied to the analysis 
of experimental data. Note that the time resolution in our model is restricted 
and the results are applicable to timescales equal or larger than $\Delta t$. 
Similarly, the temporal coarse-graining imposed by the camera frame rate in experiments 
discretizes the particle dynamics; the dynamics on time scales smaller 
than $\Delta t$ cannot be captured by interpolation since the information is lost. 

\begin{figure}[t]
\centerline{\includegraphics[width=0.4\textwidth]{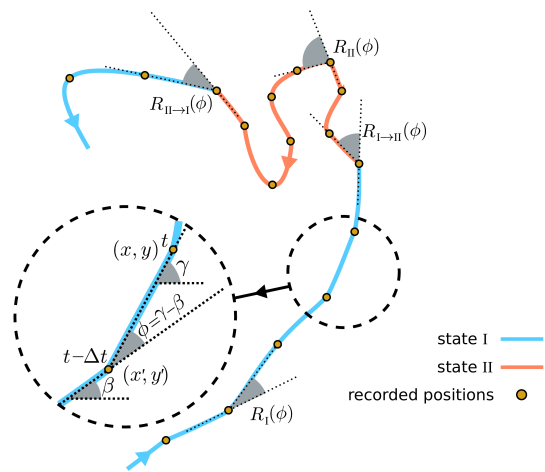}}
\caption{Sketch of a sample trajectory with two states of motion. The selected directional 
changes along the trajectory represent the four possible turning angles introduced 
in the model: the directional changes within the states (characterized by the 
turning-angle distributions $\Ri$ and $\Rf$) and the changes in the direction 
of motion at the switching events (characterized by the turning-angle distributions 
$\aif$ and $\afi$). The inset magnifies the trajectory around the successive 
timesteps $t{-}\Delta t$ and $t$. After a change $\phi\,{=}\,\gamma\,{-}\,\beta$ 
in the direction of motion at $t{-}\Delta t$, the particle arrives at position 
$(x,y)$ along the direction $\gamma$ at time $t$.}
\label{Fig:1}
\end{figure}

For generality, we assume that the particle moves with a variable instantaneous velocity. 
At each timestep, the instantaneous velocity is drawn from an arbitrary velocity distribution 
$\Fvi$ or $\Fvf$ for state $\text{I}$ or $\text{I\!I}$ of motion, respectively. The 
n-th velocity moment in the j-th state is given as $\langle v^\text{n}\rangle_{_{
\text{j}}}{=}\int_{0}^{{\infty}}\!v^\text{n}\,F\!\!_{_{\text{j}}}{(v)}\,\text{d}v$. 
For the calculation of the MSD presented in the Appendix, the relevant velocity moments 
are only the mean and the second moment, denoted with $\langle v\rangle_{_{
\text{I}}}$, $\langle v\rangle_{_{\text{I\!I}}}$, $\langle v^2\rangle_{_{\text{I}}}$, 
and $\langle v^2\rangle_{_{\text{I\!I}}}$. However, higher velocity moments also 
appear in the calculation of higher displacement moments as well as other 
transport quantities of interest.

We introduce four turning-angle distributions for the directional change $\phi$ 
of the particle between successive time points of the random walk \cite{Shaebani14,
Burov13}: $\Ri$ and $\Rf$ for turning events within the states and $\aif$ and 
$\afi$ for changing the direction of motion when switching between the states 
(see Fig.\,\ref{Fig:1}). We quantify the persistence in each state by 
\begin{eqnarray}
\nonumber
\ai&\;{=}\!\displaystyle\int_{-{\pi}}^{{\pi}}\!\!\!\!e^{i\phi}\,
R_{\text{I}}(\phi)\;\text{d}\phi, \\  
\af&\;{=}\!\displaystyle\int_{-{\pi}}^{{\pi}}\!\!\!\!e^{i\phi}\,
R_{\text{I\!I}}(\phi)\;\text{d}\phi.
\label{Eq:state_persistence}
\end{eqnarray}
In many applications, turning-angle distributions are symmetric and the persistencies 
reduce to real numbers $\ai\,{=}\,\langle\cos\phi\rangle\!_{_{R_{\text{I}}}}$ and $\af\,
{=}\,\langle\cos\phi\rangle\!_{_{R_{\text{I\!I}}}}$ in the interval $[-1,1]$. According 
to this generalized definition of the persistence, one obtains a positive $a_{_{\text{j}}}$ 
if $R_{_{\text{j}}}\!(\phi)$ is peaked around forward directions (persistent random walk). 
An isotropic $R_{_{\text{j}}}\!(\phi)$ leads to $a_{_{\text{j}}}{=}\,0$ (diffusion) 
and a distribution which is peaked around backward directions leads to a negative 
$a_{_{\text{j}}}$ (anti-persistent random walk). The extreme values $a_{_{\text{j}}}{=}\,
{+}1$ and ${-}1$ correspond to a ballistic motion and a pure localization, respectively. 
For the general case of an asymmetric $R_{_{\text{j}}}\!(\phi)$, $a_{_{\text{j}}}$ has 
a nonzero imaginary part in the absence of the left-right symmetry ($a_{_{\text{j}}}{=}
\,\langle\cos\phi\rangle\!_{_{R_{\text{j}}}}{\pm}\,i\,\langle\sin\phi\rangle
\!_{_{R_{\text{j}}}}$) which leads to a spiral trajectory \cite{Sadjadi15} (as 
observed, e.g., for the dynamics of {\it E.\,coli} near surfaces \cite{PerezIpina19}).

The particle switches stochastically between the two states with asymmetric probabilities 
$\fif$ and $\ffi$. Assuming constant transition probabilities $\fif$ and $\ffi$ leads to 
exponential residence time distributions 
\begin{eqnarray}
F\!\!_{_{\text{I}}}{(\tau)}{\sim}\,\displaystyle\text{e}^{\ln(1{-}\fif\!)\;\!\tau},\;
F\!\!_{_{\text{I\!I}}}{(\tau)}{\sim}\,\displaystyle\text{e}^{\ln(1{-}\ffi\!)\;\!\tau},
\label{Eq:residence_time_dist}
\end{eqnarray}
with the mean residence times 
\begin{eqnarray}
\langle \tau\rangle_{_{\text{I}}}{=}1{/}\fif,\;
\langle \tau\rangle_{_{\text{I\!I}}}{=}1{/}\ffi.
\label{Eq:residence_time_avg}
\end{eqnarray}
For non-exponential residence-time distributions see e.g.\ \cite{Sadjadi21,Detcheverry17}. 
For generality, we assume that each switching event is accompanied by a change in the 
direction of motion according to the turning-angle distribution $\aif$ or $\afi$. 
Similar to the persistence of each state, the persistence at each state-switching 
event can be quantified as
\begin{eqnarray}
\nonumber
\aFif&\;{=}\!\displaystyle\int_{-{\pi}}^{{\pi}}\!\!\!\!e^{i\phi}\,
\aif\;\text{d}\phi, \\  
\aFfi&\;{=}\!\displaystyle\int_{-{\pi}}^{{\pi}}\!\!\!\!e^{i\phi}\,
\afi\;\text{d}\phi,
\label{Eq:switching_persistence}
\end{eqnarray}
and assuming symmetric distributions $\aif$ and $\afi$, the above equations reduce 
to real numbers $\aFif\,{=}\,\langle\cos\phi\rangle\!_{_{R_{_{\text{I}\rightarrow
\text{I\!I}}}}}$ and $\aFfi\,{=}\,\langle\cos\phi\rangle\!_{_{R_{_{\text{I\!I}
\rightarrow\text{I}}}}}$. A turning measure $a_{_{\text{j}\rightarrow\text{j}'}}$ 
close to 1, -1, or 0 corresponds, respectively, to slightly changing, reversing, 
or randomizing the direction of motion when switching from state $\text{j}$ to 
$\text{j}'$. 

\section{Evolution of the mean square displacement}
\label{Sec:TimeEvolution}
We introduce $\Pti$ and $\Ptf$ for states $\text{I}$ and $\text{I\!I}$ of motion 
as the joint probability density functions to find the particle at time $t$ 
at position $(x,y)$ provided that the particle has reached this position along 
the direction $\gamma$ (see Fig.\,\ref{Fig:1}). We assume that a turning angle 
$\phi\,{=}\,\gamma\,{-}\,\beta$ has occurred after leaving the old position 
($x'{=}\,\xp$, $y'{=}\,\yp$) at the previous timestep $t{-}\Delta t$. The total 
probability density $P\!_t(x,y|\gamma)$ is then given by $P\!_t(x,y|\gamma)\,{=}
\,\Pti\,{+}\,\Ptf$. The stochastic process sketched in Fig.\,\ref{Fig:1} is 
described by the Master equation for the probability densities $\Pti$ and $\Ptf$:
\begin{widetext}
\begin{eqnarray}
\Ptm=\Mtran\,\Ptpm.
\label{Eq:MasterEqs}
\end{eqnarray}
\end{widetext}
Each of the master equations consists of two terms on the right hand side, 
which represent the possibility of being in each of the two states in the 
previous time step. The change in the direction of motion $\phi{=}\gamma
{-}\beta$ is randomly chosen from the four turning-angle distributions. 
Here, the velocity and turning-angle distributions are independent but 
they can be correlated in general \cite{Shaebani20,Cherstvy18}. Also, 
successive velocities are assumed to be uncorrelated for simplicity 
but they can be correlated in general. In such a case, $F(v)$ can be 
replaced with a velocity-change distribution similar to $R(\gamma\,{-}
\,\beta)$. This would lead to a convolution form after Fourier transform, 
which cannot be solved in the general form. However, it is possible to 
solve the problem for at least some explicit forms of the velocity-change 
distribution. Using the Fourier transform of the probability density function 
\begin{equation}
\mathcal{\widetilde{P}}_t({\bm k}|m)\,{=}\!\!\int\limits_{-\pi}^{\;\;\pi}
\!\!\text{d}\gamma\,e^{im\gamma} \!\!\int \!\! \text{d}y \!\!\int\!\!
\text{d}x\,\,e^{i\boldsymbol{k}\cdot\boldsymbol r} P\!_t(x,y|\gamma),
\label{Eq:P_Fourier}
\end{equation}
the displacement moments can be extracted as
\begin{equation}
\displaystyle\langle x^{a} y^{b} \rangle(t)= (-i)^{a+b} \frac{\partial^{a+b} 
\mathcal{\widetilde{P}}_t(k_{x},k_{y}|m=0)}{\partial k_{x}^{a}\partial k_{y}^{b}} 
\Big|_{(k_{x},k_{y})=(0,0)}.
\label{Eq:Moments}
\end{equation}
For example, using the polar representation of $\boldsymbol{k}$ as $(k,\alpha)$, 
the $x$ component of the MSD can be calculated as 
\begin{equation}
\displaystyle\langle x^{2} \rangle(t)=(-i)^{2} \frac{\partial^{2} 
\mathcal{\widetilde{P}}_{\!_{t}}(k,\alpha{=}0|m{=}0)}{\partial k^{2}} \Big|_{k{=}0}.
\label{Eq:MSD_x}
\end{equation} 
We present a Fourier-z-transform technique in the Appendix to obtain analytical 
expressions for arbitrary displacement moments for the stochastic process 
described by the master equations (\ref{Eq:MasterEqs}); see also \cite{Sadjadi15,
Sadjadi08}. The recipe to obtain an arbitrary displacement moment is provided 
and the calculations are shown in detail to obtain an exact expression for the 
time evolution of the MSD, which is of broad interest. The formalism can be 
extended to extract other transport quantities such as the first-passage 
properties. 

The initial probabilities $q_{_0}^{\text{I}}$ and $q_{_0}^{\text{I\!I}}$ 
of starting in state $\text{I}$ or $\text{I\!I}$ influence the short-time 
dynamics but after some time the probabilities $q_{_t}^{\text{I}}$ and 
$q_{_t}^{\text{I\!I}}$ eventually converge to their steady state values 
$q_{\text{steady}}^{\text{I}}$ and $q_{\text{steady}}^{\text{I\!I}}$, 
that are not only independent of the initial probabilities 
($q_{_0}^{\text{I}},\,q_{_0}^{\text{I\!I}}$) but also independent 
of time. Note that the process of being in each state is different 
from the original process defined by the particle positions and 
directions. Thus, the steady state of $q_{_t}^{\text{I}}$ and 
$q_{_t}^{\text{I\!I}}$ differs from the long-time diffusive 
dynamics regime of $P\!\!_t(x,y|\gamma)$. To estimate the timescale 
for reaching the steady state, the sequence of being in state $\text{I}$ 
or $\text{I\!I}$ can be considered as a restricted Markov chain with 
transition probabilities $\fif$ and $\ffi$. By solving 
\begin{equation}
\left(\!\!\begin{array}{cc}
q_{_t}^{\text{I}} \\
\\
q_{_t}^{\text{I\!I}} 
\end{array}\!\!\right){=}\left(\!\!\begin{array}{cc}
1{-}\fif\,\,\,\,\,\,\ffi\,\,\, \\
\\
\,\,\,\fif\,\,\,\,\,\,1{-}\ffi 
\end{array}\!\!\right)^{\!t} \left(\!\!\begin{array}{cc}
q_{_0}^{\text{I}} \\
\\
q_{_0}^{\text{I\!I}} 
\end{array}\!\!\right),
\label{Eq:Markov_Chain}
\end{equation}
it can be verified that the time evolution of the restricted Markov chain follows 
\begin{eqnarray}\nonumber
q_{_t}^{\text{I}}\,&{=}\,\displaystyle\frac{\ffi}{\fif{+}\ffi}+
\displaystyle\frac{(1{-}\ffi{-}\fif)^t}{\fif{+}\ffi}(\fif\,
q_{_0}^{\text{I}}{-}\ffi\,q_{_0}^{\text{I\!I}}), \\ \nonumber
q_{_t}^{\text{I\!I}}\,&{=}\,\displaystyle\frac{\fif}{\fif{+}\ffi}+
\displaystyle\frac{(1{-}\ffi{-}\fif)^t}{\fif{+}\ffi}(\ffi\,
q_{_0}^{\text{I\!I}}{-}\fif\,q_{_0}^{\text{I}}). \\
\label{Eq:Markov_Evolve}
\end{eqnarray}
Thus, the restricted Markov process described by the transitions between 
state $\text{I}$ and $\text{I\!I}$ exponentially approaches the steady 
state probabilities 
\begin{eqnarray}
q_{\text{steady}}^{\text{I}}=\displaystyle\frac{\ffi}{\fif{+}\ffi},\;
q_{\text{steady}}^{\text{I\!I}}=\displaystyle\frac{\fif}{\fif{+}\ffi}.
\label{Eq:Markov_Steady}
\end{eqnarray}
If the process initially starts with the steady state probabilities, i.e.\ 
$q_{_0}^{\text{I}}\,{=}\,q_{\text{steady}}^{\text{I}}$ and $q_{_0}^{
\text{I\!I}}\,{=}\,q_{\text{steady}}^{\text{I\!I}}$, the last 
parentheses on the right hand side of Eqs.\,(\ref{Eq:Markov_Evolve}) 
will be zero and the restricted Markov process is immediately in the steady 
state. Similarly, for the specific choice of $\fif\,{+}\,\ffi\,{=}\,1$, the 
exponential terms on the right hand side of Eqs.\,(\ref{Eq:Markov_Evolve}) 
will be zero and again the the restricted Markov process is immediately in 
the steady state. The characteristic time to exponentially approach the 
steady state can be obtained from Eqs.\,(\ref{Eq:Markov_Evolve}) as
\begin{equation}
\ts=\displaystyle\frac{-1}{\ln|1{-}\fif{-}\ffi|}\,. 
\label{Eq:ts}
\end{equation}

By choosing steady state initial conditions--- i.e. $q_{_0}^{\text{I}}
\,{=}\,q_{\text{steady}}^{\text{I}}$ and $q_{_0}^{\text{I\!I}}\,{=}\,
q_{\text{steady}}^{\text{I\!I}}$---, we exclude the role of the initial 
conditions and reduce the complexity of the short-time dynamics. For 
this choice and an isotropic initial orientation, after some calculations 
(see Appendix) we obtain the following exact expression for the time 
evolution of the MSD 
\begin{eqnarray}
\langle r^2\rangle(t)=\mathcal{A}+\mathcal{B}\,t+\mathcal{C}\,
\text{e}^{-t{/}\tcp}+\mathcal{D}\,\text{e}^{-t{/}\tcm},
\label{Eq:Full_MSD}
\end{eqnarray}
with characteristic times  
\begin{widetext}
\begin{equation}
t_{\text{c}_{\pm}}{=}\displaystyle\frac{-1}{\ln\Bigg|\frac{\ai(1{-}\fif)\,{+}
\,\af(1{-}\ffi){\pm}\sqrt{\big(\ai(1{-}\fif)\,{-}\,\af(1{-}\ffi)\big)^2\!{+}\,
4\,\aFif\,\aFfi\,\fif\,\ffi}}{2}\Bigg|}, 
\label{Eq:tc}
\end{equation}
\end{widetext}
which reduces to $t_{\text{c}}{=}\frac{-1}{\ln|a|}$ for a single-state persistent 
random walk (i.e.\ for $\fif{=}0$ and $\ffi{=}1$) with persistence $a$. The 
time-independent term $\mathcal{A}$ and the prefactors $\mathcal{B}$, 
$\mathcal{C}$ and $\mathcal{D}$ are functions of the persistencies 
($\ai$,\,$\af$,\,$\aFif$,\,$\aFfi$), the switching probabilities 
($\fif$,\,$\ffi$), and the first two velocity moments 
($\langle v\rangle_{_{\text{I}}}$,\,$\langle v\rangle_{_{\text{I\!I}}}$,\,$\langle 
v^2\rangle_{_{\text{I}}}$,\,$\langle v^2\rangle_{_{\text{I\!I}}}$); see 
Appendix. Equation\,(\ref{Eq:Full_MSD}) shows that the MSD consists 
of exponentially-decaying terms with $t$, a time-independent term, 
and a term which grows linearly with $t$. The short-time dynamics 
is mainly controlled by the exponentially-decaying and time-independent 
terms, whereas the linear term dominates at long times. Note 
that only the first Fourier mode of each turning-angle distribution (i.e.\ 
$\langle \cos\phi\rangle$) appears in the calculation of the MSD and 
the overall form the distribution plays no role. Higher Fourier modes of the 
turning-angle distribution appear in the calculation of the higher 
displacement moments. For example, $\langle \cos(2\phi) \rangle$ appears in 
the expression for $\langle r^4 \rangle(t)$ \cite{Sadjadi15}. Although 
Eq.\,(\ref{Eq:Full_MSD}) is a profitable expression to compare with the 
experimental data, we advise (especially for non-steady state initial 
conditions) to insert the parameter values into the much shorter form 
of the MSD before the inverse z-transform, i.e.\ $\langle r^2 \rangle(z)$ 
given in Eq.\,(\ref{Eq:MSDz}), and then use Mathematica or any other 
software to apply the inverse z-transform and obtain the analytical 
form of the MSD as a function of $t$.

\begin{figure}[t]
\centerline{\includegraphics[width=0.47\textwidth]{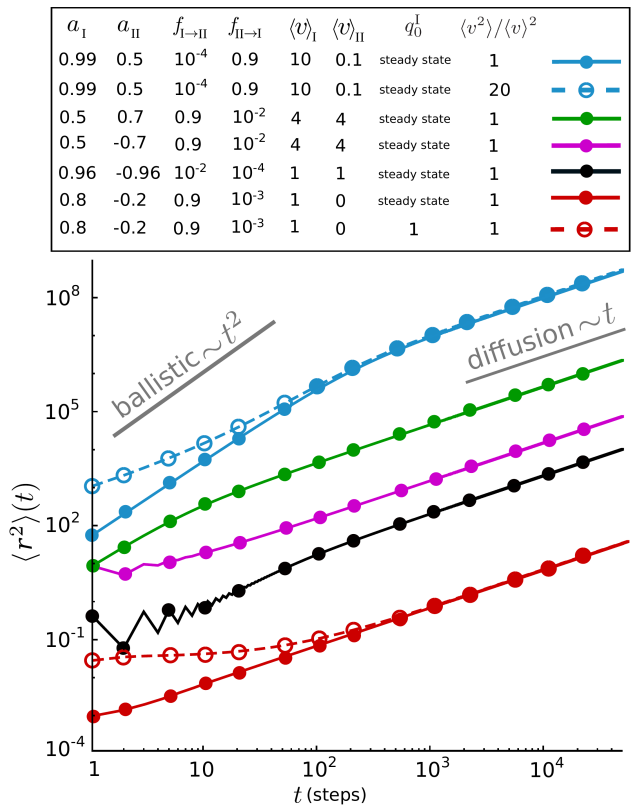}}
\caption{Time evolution of the MSD in log-log scale. The symbols denote 
simulation results and the lines correspond to analytical predictions 
via Eq.\,(\ref{Eq:Full_MSD}). A rich variety of dynamical regimes 
of anomalous motion can be observed at short and intermediate time 
scales. The velocity of each state is constant except for the blue (upper) 
dashed curve, which is obtained for broad uniform velocity distributions 
(with $\langle v^2\rangle\,{=}\,20\,\langle v\rangle^2$ in both 
states). All the MSD profiles belong to steady state initial 
conditions except for the red (lower) dashed curve which is obtained for the 
initial condition $q_{_0}^{\text{I}}\,{=}\,1$, i.e.\ starting the 
motion in state $\text{I}$. In all plots we have chosen $\aFif\,{=}\,
\af$ and $\aFfi\,{=}\,\ai$ for the directional changes at the switching 
events. An ensemble of $10^5$ realizations has been considered for the 
simulations and the four turning-angle distributions are uniformly 
distributed around the zero change in the turning angle.}
\label{Fig:2}
\end{figure}

Figure\,\ref{Fig:2} shows the time evolution of MSD over a wide range of 
time scales. Various types of anomalous diffusion (i.e.\ MSD $\langle r^2\rangle(t)$ 
not proportional to $t$) can be observed upon varying 
the key parameters. For simplicity, we have presented the results for $\aFif\,
{=}\,\af$ and $\aFfi\,{=}\,\ai$ at the switching events, a constant velocity 
in each state, and steady state initial conditions (solid lines and symbols), 
unless specified otherwise. For visibility, different velocity values are used 
to separate the curves from each other. The shape of the MSD profile strongly 
depends on the choice of persistence parameters and switching probabilities. 
For a combination of two persistent random walks, the crossover from the 
initial superdiffusive to the asymptotic diffusive dynamics can be delayed 
by increasing the persistencies or the residence time in the more persistent 
state. A mixture of a persistent and a slightly antipersistent walk results 
in a subdiffusive dynamics at short times. However, by choosing a strongly 
antipersistent state, an oscillatory dynamics at short timescales emerges 
\cite{Tierno12,Tierno16}. In some parameter regimes, the exponential terms 
of the MSD rapidly decay and time-independent terms develop a plateau regime 
over intermediate timescales; see, e.g., the red (lower) dashed curve obtained 
by choosing $q_{_0}^{\text{I}}\,{=}\,1$ instead of steady state initial 
conditions. The initial conditions influence the time-independent and 
exponentially-decaying terms of the MSD and diversify the anomalous diffusion 
on short time scales. However, the term which grows linearly with $t$ is 
independent of the initial conditions, thus, the profiles for different 
initial conditions eventually merge at long times when the crossover to 
asymptotic diffusive dynamics occurs. 

We have presented the formalism in the Appendix for an arbitrary choice 
of the initial condition until Eq.\,(\ref{Eq:MSDz}) for the MSD in the 
z-space but then solved the inverse z-transform problem for the specific 
choice of steady state initial conditions. Thus, by inserting any desired 
initial condition into Eq.\,(\ref{Eq:MSDz}) and calculating the inverse 
z-transform, the time evolution of MSD can be obtained (as we did for the 
red (lower) dashed curve in Fig.\,\ref{Fig:2}). 

To see how the velocity variations influence the MSD profile, we present 
an example of variable velocities in Fig.\,\ref{Fig:2} (blue (upper) dashed curve): 
A wide velocity distribution in each state is chosen ($\langle v^2\rangle_{_{
\text{I}}}\,{=}\,20\,\langle v\rangle^2_{_{\text{I}}}$ and $\langle v^2
\rangle_{_{\text{I\!I}}}\,{=}\,20\,\langle v\rangle^2_{_{\text{I\!I}}}$). 
It can be seen that the broadening of the velocity distributions unexpectedly 
pushes the initial slope towards the diffusion line by increasing the 
role of the linear term of the MSD.

\begin{figure*}[t]
\centerline{\includegraphics[width=0.99\textwidth]{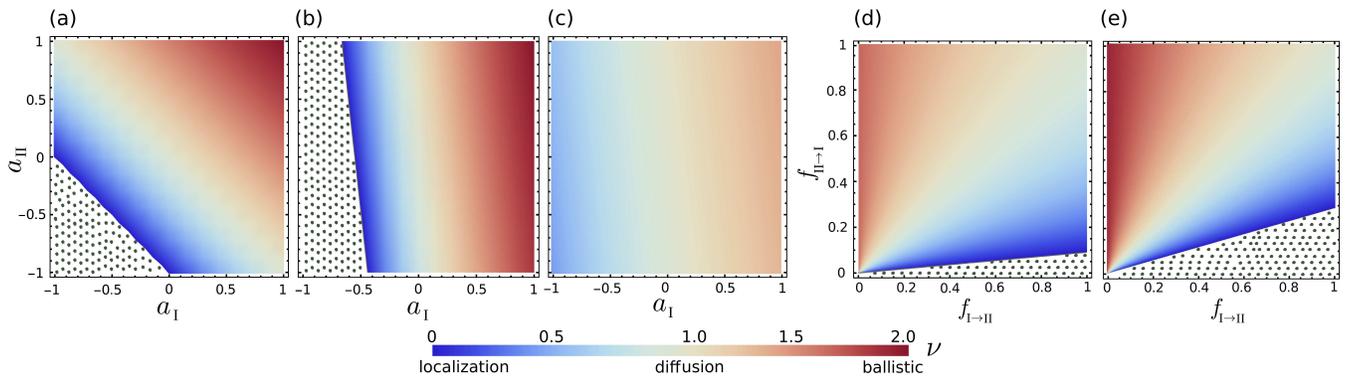}}
\caption{Phase diagrams of the initial anomalous exponent according 
to Eq.\,(\ref{Eq:Exponent2}). The color intensity reflects the magnitude 
of $\nu$, with red (blue) meaning superdiffusion (subdiffusion). 
The dotted regions denote oscillatory subdomains. We consider 
$\aFif\,{=}\,\af$ and $\aFfi\,{=}\,\ai$ in all panels and choose 
a constant velocity ($\langle v\rangle\,{=}\,\langle v^2\rangle\,{=}
\,1$) in both states, except for panel (c). (a,\,b) $\nu$ in $(\ai,\af)$ 
plane for (a) $\fif\,{=}\,\ffi\,{=}\,0.5$ and (b) $\fif\,{=}\,0.1$, 
$\ffi\,{=}\,0.9$. By the asymmetric choices of $\fif$ and $\ffi$, 
$\nu$ is influenced stronger by the persistence of the state with 
a longer mean residence time. (c) Similar parameter values as in panel 
(b) but with broad velocity distributions ($\langle v\rangle\,{=}\,1$ 
and $\langle v^2\rangle\,{=}\,3$ in both states). Broader velocity 
distributions push the initial slope of the MSD towards $\nu\,{=}
\,1$ (diffusion). (d,\,e) $\nu$ in $(\fif,\ffi)$ plane for a 
combination of persistent and antipersistent motions characterized 
by (d) $\ai\,{=}\,0.6$, $\af\,{=}\,-0.6$ and (e) $\ai\,{=}\,0.9$, 
$\af\,{=}\,-0.9$. $\nu$ is enhanced by the switching probabilities 
that lead to a longer stay in the persistent mode.}
\label{Fig:3}
\end{figure*}

To confirm the validity of the analytical predictions, we perform extensive 
Monte Carlo simulations of the stochastic process. We consider a 2D 
persistent random walk with two different modes of motion and allow the 
walker to spontaneously change the mode of motion at each timestep according 
to given asymmetric transition probabilities. By choosing an arbitrary length 
unit, the step size has been varied within the range [0.01,\,100] depending 
on the choice of the velocity distributions. Periodic boundary conditions are 
imposed and the results are shown for the system size $L\,{=}\,10^5$. The 
walker starts at the center of the simulation box and the initial orientation 
of motion is randomly drawn from an isotropic distribution. For the velocity 
and turning-angle distributions, we have chosen uniform distributions which 
are symmetric around the mean velocity of each state or the turning angle 
$\phi\,{=}\,0$, respectively. However, we note that choosing other arbitrary 
distributions with the same first two velocity moments and mean persistence 
$\langle\cos\phi\rangle$ lead to exactly the same MSD profile (though, higher 
displacement moments would differ from those of the uniform distribution choice). 
Figure\,\ref{Fig:2} shows the simulation results, averaged over an ensemble 
of $10^5$ realizations. The simulation results agree perfectly with the 
analytical predictions.

\section{Initial anomalous exponent}
\label{Sec:InitialExponent}
The transient dynamics is of particular interest as the time window 
of experiments is practically limited. To characterize and compare 
the initial growth rate of the MSD profiles, one can assign an initial 
anomalous exponent to each MSD curve by fitting it to a power-law 
$\langle r^2 \rangle(t){\sim}t^{\nu}$. To this aim, we obtain the 
MSD of the first two points along the curve from Eq.\,(\ref{Eq:Full_MSD}) 
as $\langle r^2\rangle(t{=}1)\,{=}\,\mathcal{A}\,{+}\,\mathcal{B}
\,{+}\,\mathcal{C}\,\text{e}^{-1{/}\tcp}+\mathcal{D}\,\text{e}^{-1{/}
\tcm}$ and $\langle r^2\rangle(t{=}2)\,{=}\,\mathcal{A}\,{+}\,2\,
\mathcal{B}\,{+}\,\mathcal{C}\,\text{e}^{-2{/}\tcp}+\mathcal{D}\,
\text{e}^{-2{/}\tcm}$. The power-law fit passes both of these 
points leading to $\ln\,\langle r^2\rangle(t{=}2)-\ln\,\langle 
r^2\rangle(t{=}1)=\nu\,(\ln 2-\ln 1)$, from which the initial 
anomalous exponent can be obtained as
\begin{equation}
\nu=\ln\Big(\displaystyle\frac{\langle r^2\rangle(t{=}2)}{\langle 
r^2\rangle(t{=}1)}\Big){\Big/}\ln 2.
\label{Eq:Exponent1}
\end{equation}
By replacing the MSD from Eq.\,(\ref{Eq:Full_MSD}) and after some 
algebra, we derive the initial anomalous exponent
\begin{widetext}
\begin{equation}
\nu{=}1{+}\ln\Big(\!1{+}\frac{\fif\,\ffi\,\langle v
\rangle_{_{\text{I}}}\,\langle v\rangle_{_{\text{I\!I}}}\,(\aFif\,
{+}\,\aFfi)\,{+}\,\ai\,\langle v\rangle^2_{_{\text{I}}}\,(1\,{-}\,
\fif)\,\ffi\,{+}\,\af\,\langle v\rangle^2_{_{\text{I\!I}}}\,\fif\,
(1\,{-}\,\ffi)}{\fif\,\langle v^2\rangle_{_{\text{I\!I}}}\,{+}\,
\ffi\,\langle v^2\rangle_{_{\text{I}}}}\Big){\Big/}\ln 2.
\label{Eq:Exponent2}
\end{equation}
\end{widetext}
For a single-state active motion (i.e.\ $\fif{=}0$ and $\ffi{=}1$) 
with persistence $a$ and constant velocity, the above equation 
reduces to $\nu\,{=}\,1\!+\frac{\ln(1{+}a)}{\ln2}$ \cite{Shaebani14}. 
Since all four persistence parameters appear linearly and with similar 
prefactors, we choose $\aFif\,{=}\,\af$ and $\aFfi\,{=}\,\ai$ at the 
switching events for simplicity. In the phase diagrams presented in 
Fig.\,\ref{Fig:3}, we show how $\nu$ depends on the remaining key 
parameters. $\nu$ ranges from 2 for ballistic motion to 1 for diffusion 
and 0 for zero net displacement. The onset of oscillatory dynamics for 
a strongly antipersistent random walk can be identified by setting 
$\nu\,{=}\,0$. Figure\,\ref{Fig:3}(a) shows that $\nu$ varies symmetrically 
in ($\ai$,\,$\af$) plane for symmetric transitions between the states, 
i.e.\ for $\fif\,{=}\,\ffi$. However, for an asymmetric choice of 
the switching probabilities $\fif\,{<}\,\ffi$, $\nu$ is more sensitive 
to the persistence of state I, which has a longer mean residence time 
according to Eq.\,(\ref{Eq:residence_time_avg}); see panel (b). Using 
broad velocity distributions as in panel (c) increases the denominator 
in Eq.\,(\ref{Eq:Exponent2}) and results in considerably smaller 
anomalous exponents, which is consistent with the short-time behavior 
of the MSD in Fig.\,\ref{Fig:2} (blue (upper) dashed curve vs blue (upper) 
solid curve). For a mixture of persistent ($\ai\,{>}\,0$) and antipersistent 
($\af\,{<}\,0$) states in panel (d), the combination of switching probabilities 
that increases the residence time in the persistent state (i.e.\ a smaller 
$\fif$ and a larger $\ffi$) enhances the anomalous exponent. The effect 
becomes stronger with increasing the magnitude of the persistencies 
$\ai$ and $\af$ in panel (e). 

\begin{figure}[t]
\centerline{\includegraphics[width=0.46\textwidth]{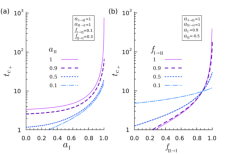}}
\caption{Characteristic time $\tcp$ via Eq.\,(\ref{Eq:tc}) in terms of (a) 
$\ai$ and (b) $\ffi$. The switching persistencies are chosen as $\aFif\,{=}
\,\aFfi\,{=}\,1$ in both panels. The crossover time grows by several orders 
of magnitude by increasing the persistence of the states or increasing the 
residence time in the more persistent mode. In panel (a), the results for 
several values of $\af$ are shown at a given set of $\fif$ and $\ffi$ 
parameters. In panel (b), $\ai$ and $\af$ are fixed and $\tcp$ is shown 
vs $\ffi$ for several values of $\fif$.}
\label{Fig:4}
\end{figure}

\begin{figure*}[t]
\centerline{\includegraphics[width=0.85\textwidth]{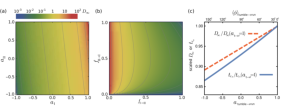}}
\caption{(a),(b) $D\!_{_\infty}$ in $(\ai,\af)$ and $(\fif,\ffi)$ planes, 
scaled by $v^2\Delta t$, for a constant velocity $v_{_{\text{I}}}{=}
v_{_{\text{I\!I}}}{=}v$ and the parameter values (unless varied): $\fif{=}
0.1$, $\ffi{=}0.9$, $\ai{=}0.9$, $\af{=}-0.9$, $\aFfi{=}\ai$, $\aFif{=}
\af$. By the chosen asymmetric switching probabilities in panel (a), 
the walker spends longer times in state I and $D\!_{_\infty}$ is 
more sensitive to $\ai$. Panel (b) shows that tuning the residence time 
in each state via $\fif$ and $\ffi$ in a combination of persistent and 
antipersistent motions dramatically influences $D\!_{_\infty}$. (c) 
Variations of $D\!_{_\infty}$ and $\tcp$ in a run-and-tumble process 
in terms of the mean tumble-to-run turning angle $\langle\phi
\rangle_{\text{tumble}{\rightarrow}\text{run}}$ (correspondingly 
$a_{_{\text{tumble}\rightarrow\text{run}}}{=}\langle\cos\phi
\rangle_{_{\text{tumble}\rightarrow\text{run}}}$). Other parameter 
values: $f\!_{_{\text{run}\rightarrow\text{tumble}}}{=}f\!_{_{\text{tumble}\rightarrow
\text{run}}}{=}0.1$, $a_{_{\text{run}}}{=}0.9$, $a_{_{\text{tumble}}}{=}0$, 
$a_{_{\text{run}\rightarrow\text{tumble}}}{=}1$, and constant velocities 
$v_{_{\text{run}}}{=}2\,v_{_{\text{tumble}}}$.}
\label{Fig:5}
\end{figure*}

\section{Crossover time to asymptotic normal diffusion}
\label{Sec:CrossoverTime}
Asymptotically the stochastic process considered here is described by 
normal diffusion (i.e.\ a non-persistent motion) since it gradually 
loses its memory of the initial direction and state of motion, and 
the trajectory eventually gets randomized. It can be seen from 
Eq.\,(\ref{Eq:Full_MSD}) that while the contribution of the term 
linear in $t$ dominates at large times, the exponential terms decay. 
To estimate the crossover time to the asymptotic diffusive regime one 
can, e.g., measure the instantaneous anomalous exponent (similar to the 
procedure explained in the previous section for the calculation of the 
initial anomalous exponent $\nu$) and follow its convergence towards 1. 
Alternatively, the characteristic times of the exponentially-decaying 
terms of the MSD (i.e.\ $\tcp$ and $\tcm$) reflect the timescale to 
approach the long-term dynamics. We follow the later choice and use 
Eq.\,(\ref{Eq:tc}) to show how the crossover time depends on the 
key parameters of the particle dynamics. Both $\tcp$ and $\tcm$ depend 
on the four persistencies $\ai$, $\af$, $\aFif$, and $\aFfi$ and the 
switching probabilities $\fif$ and $\ffi$ but are independent of the 
initial conditions and the velocity distributions.

To reduce the degrees of freedom, we set $\aFif=\aFfi=1$ 
(corresponding to move straight forward at the switching events). 
Figure\,\ref{Fig:4} shows the behavior of $\tcp$, as an example. 
Similar results can be deduced for $\tcm$ as well. The characteristic 
time can vary by several orders of magnitude upon changing the remaining 
control parameters. For a given set of ($\fif$,\,$\ffi$) and a combination 
of two persistent random walks, it is shown in Fig.\,\ref{Fig:4}(a) 
that $\tcp$ grows with increasing $\ai$ and $\af$. For a mixture 
of persistent $\ai$ and antipersistent $\af$ states, Fig.\,\ref{Fig:4}(b) 
reveals that the switching probabilities that lead to a longer 
residence in the persistent mode (i.e.\ a larger $\ffi$ or a smaller 
$\fif$) enhance $\tcp$ even by orders of magnitude. We note that 
the velocity moments may also influence the crossover time through 
the prefactors $\mathcal{C}$ and $\mathcal{D}$ in Eq.\,(\ref{Eq:Full_MSD}). \\

\section{Asymptotic diffusion constant}
\label{Sec:AsymptoticD}
According to Eq.\,(\ref{Eq:Full_MSD}), the exponential terms of the MSD 
gradually decay and the time-independent term also becomes negligible 
compared to the linear term at long times. As the linear term eventually 
dominates, the process is asymptotically diffusive and the MSD follows 
$\langle r^2 \rangle(t{\rightarrow}\infty)\,{\sim}\,\mathcal{B}\,t$. 
By writing the MSD in the diffusion regime as $\langle r^2 \rangle(t
{\rightarrow}\infty)\,{=}\,2\,d\,D\!_{_\infty}t$ (with $d\,{=}\,2$ 
being the dimension of the system), the long-term diffusion coefficient 
can be deduced as
\begin{widetext}
\begin{equation}
\begin{aligned}
D\!_{_\infty}{=}&\frac{\Delta t}{4}\frac{\ffi(e_{_2}{-}1)\Vdoi{+}
\fif(e_{_1}{-}1)\Vdof{-}2\,e_{_4}\vi\vf{-}\ffi\big((e_{_2}{-}1)
e_{_1}{-}e_{_3}\big)\Delta\,v_{_{\text{I}}}{-}\fif\big((e_{_1}{-}1)
e_{_2}{-}e_{_3}\big)\Delta\,v_{_{\text{I\!I}}}}{(\fif{+}\ffi)\,
\Big(e_{_3}\,{-}\,(e_{_1}{-}1)(e_{_2}{-}1)\Big)},
\end{aligned}
\label{Eq:Dasymp}
\end{equation}
\end{widetext}
with $\Delta \vj{=}\langle v^2\rangle_{_{\text{j}}}{-}2\vjm^2$, $e_{_1}
{=}(1{-}\fif)\ai$, $e_{_2}{=}(1{-}\ffi)\af$, $e_{_3}{=}\fif\ffi\aFif\aFfi$, 
and $e_{_4}{=}\fif\ffi(\aFif{+}\aFfi)$. For a single-state persistent random 
walk (i.e.\ $\fif{=}0$ and $\ffi{=}1$) with persistency $a$, the above equation 
reduces to $D\!_{_\infty}{=}\frac{\Delta t}{4}\big(\langle v^2\rangle{+}
\frac{2\,a}{1{-}a}\langle v\rangle^2\big)$ \cite{Shaebani14}. 

To visualize $D\!_{_\infty}$ in terms of the key parameters, we consider a 
process in which $\aFfi{=}\ai$ and $\aFif{=}\af$ at the switching events. 
As shown in Figs.\,\ref{Fig:5}(a),(b), $D\!_{_\infty}$ varies 
by several orders of magnitude by changing the key parameters ($\ai$,\,$\af$) 
or ($\fif$,\,$\ffi$). Equation\,(\ref{Eq:Dasymp}) describes $D\!_{_\infty}$ 
for any arbitrary combination of the stochastic processes. For instance, 
for a simple combination of diffusion (with constant $D_{_{\text{I}}}$) 
and waiting, Eq.\,(\ref{Eq:Dasymp}) reduces to $D\!_{_\infty}{=}
D_{_{\text{I}}}\displaystyle\frac{\ffi}{\fif{+}\ffi}$, which was originally 
shown by Lennard-Jones for surface diffusion with traps \cite{Lennard-Jones32}. 
$D\!_{_\infty}$ is independent of the initial conditions $q_{_0}^{\text{I}}$ 
and $q_{_0}^{\text{I\!I}}$, implying that the history of the process is 
only carried by exponential and time-independent terms of the MSD that 
are negligible at long times as the linear term eventually dominates. 

\begin{figure}[b]
\centerline{\includegraphics[width=0.45\textwidth]{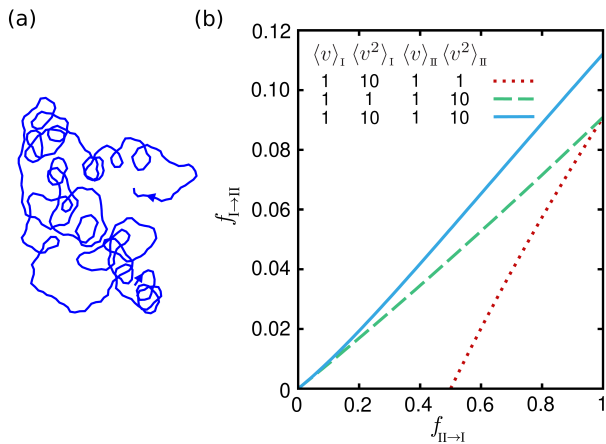}}
\caption{(a) Typical trajectory of a single-state persistent walker 
with an asymmetric turning-angle distribution uniformly distributed 
over the range $\phi\,{\in}\,[{-}\frac{\pi}{6}, \frac{2\pi}{6}]$.
The asymmetry of the turning angles leads to a trajectory with 
frequent clockwise spirals. The resulting persistence parameter 
has real and imaginary parts $a_\text{R}\,{\simeq}0.9$ and $a_\text{I}
\,{\simeq}0.2$, respectively. (b) Subset of optimal switching 
probabilities $\fif^{\,\,\text{opt}}$ and $\ffi^{\,\,\text{opt}}$ 
that maximize the asymptotic diffusion constant according to 
Eqs.\,(\ref{Eq:Dderivative}) and (\ref{Eq:Doptimized}) in the 
($\fif,\ffi$) plane for various choices of the velocity moments 
in a combination of ballistic and diffusive processes.}
\label{Fig:6}
\end{figure}

\section{Applications and special cases}
\label{Sec:Applications}
The broad applicability of our formalism allows generic predictions about 
the dynamics of various systems. In this section we present a few 
applications and the reduced form of the general analytical expressions 
for a couple of specific choices for the states of motion.

{\it Bacterial dynamics}--- Bacterial species that swim by the rotation 
of flagella experience an alternating sequence of run and tumble phases. 
An abrupt directional change often occurs when switching back from tumble 
to run phase \cite{Berg04,Najafi18}, which is caused by the torque exerted 
on the cell body during the reformation of the bundle \cite{Darnton07}. It 
is hypothesized that the bacteria benefit from this feature to slow their 
spreading and explore the local environment more precisely. Since in our 
model the statistics of the turning angles at the switching events are chosen 
to be independent from the turning angles within the states in general, we 
can directly check how directional changes at the switching events influence 
the crossover time to the long-term diffusive dynamics and the asymptotic 
diffusion coefficient. Figure\,\ref{Fig:5}(c) shows that increasing the 
mean directional change at switching from tumble to run, $\langle\phi
\rangle_{\text{tumble}{\rightarrow}\text{run}}$, helps the bacteria 
to randomize their path: A stronger kick (i.e.\ a larger turning angle 
$\langle\phi\rangle_{\text{tumble}{\rightarrow}\text{run}}$) can reduce 
the crossover time $t_{\text{c}_{+}}$ and the diffusion coefficient 
$D\!_{_\infty}$ by more than $10\%$ for the given set of parameters 
in Fig.\,\ref{Fig:5}(c); for a higher run persistence 
$a_{_{\text{run}}}{=}\,0.96$, the percentage of reduction can even 
exceed $25\%$.  

{\it Spiral trajectories}--- The turning-angle distributions $R_{\text{I}}(\phi)$, 
$R_{\text{I\!I}}(\phi)$, $\aif$ and $\afi$ can be asymmetric in general. For 
any asymmetric distribution $R(\phi)$, the persistence introduced 
in Eq.\,(\ref{Eq:state_persistence}) or (\ref{Eq:switching_persistence}) 
has a real part $a_{_\text{R}}{=}\dphi\,\cos\phi\,R(\phi){=}\langle
\cos\phi\rangle$ and a nonzero imaginary part $a_{_\text{Im}}{=}\dphi\,
\sin\phi\,R(\phi){=}\langle\sin\phi\rangle$. If we consider a single-state 
2D motion for simplicity, an asymmetric $R(\phi)$ means that the left-right 
symmetry does not hold and the particle turns more frequently either to the 
right or to the left direction, leading to the emergence of clockwise or 
anti-clockwise spiral trajectories. For example, motion with a uniform 
distribution $R(\phi)\,{=}\,\frac{1}{\pi{/}2}$ but over an asymmetric 
range $[{-}\frac{\pi}{6}, \frac{2\pi}{6}]$ of $\phi$ corresponds to a 
trajectory with frequent clockwise spirals; see Fig.~\ref{Fig:6}(a). 
Following the analytical approach presented in the Appendix, one can 
choose the conditions for a single state (and also set $\alpha{=}m{=}0$ 
in Eq.\,(\ref{Eq:Q_i_c})) to obtain the expansion coefficient that is 
required to extract the MSD. Because of the asymmetry of the 
turning-angle distribution, $\mathcal{\widetilde{R}}(m{=}1){=}
a_{_\text{R}}{+}i\,a_{_\text{Im}}$ and $\mathcal{\widetilde{R}}(m{=}
{-}1){=}a_{_\text{R}}{-}i\,a_{_\text{Im}}$, i.e.\ $\mathcal{\widetilde{R}}
(m{=}1){\neq}\mathcal{\widetilde{R}}(m{=}{-}1)$; see Eq.\,(\ref{Eq:Fourier_Distributions}). 
By inserting these quantities into Eq.\,(\ref{Eq:Q_i_c}) and following 
the rest of the procedure, the MSD can be obtained. From the linear 
term of the MSD in $t$, the asymptotic diffusion constant can be 
deduced as
\begin{equation}
D\!_{_\infty}{=}\frac{1}{4}\,v^2\!\Delta t\,\frac{a_{_\text{R}}(1{-}
a_{_\text{R}}){-}a_{_\text{Im}}^2}{(1{-}a_{_\text{R}}\!)^2{+}
a_{_\text{Im}}^2},
\label{Eq:DasympAsymmetric}
\end{equation}
where we choose a constant velocity $v$ for simplicity. It can be seen that 
the asymmetric contribution reduces the asymptotic diffusion coefficient. 
If we denote the diffusion coefficient of a diffusion process ($a_{_\text{R}}
{=}\,a_{_\text{Im}}{=}\,0$) with $D\!_{_0}{=}\frac{1}{4}\,v^2\!\Delta t$, 
$D\!_{_\infty}$ can be even smaller than $D\!_{_0}$ despite of having 
a positive real part of the persistence ($a_{_\text{R}}\,{>}\,0$). 
A constraint for a pure localization ($D\!_{_\infty}{=}\,0$) can be 
obtained as $a_{_\text{R}}^2\,{+}\,a_{_\text{Im}}^2\,{=}\,a_{_\text{R}}$.

{\it Run-and-tumble dynamics}--- A subclass of two-state processes of particular 
interest is a combination of fast and slow dynamics, described by the so-called 
{\it run-and-tumble} models \cite{Angelani09,Elgeti15,Thiel12,Angelani13}. 
The modeling of such processes has been often limited either to extract the 
long-term dynamics of the particle or to simplify the states with stochastic 
processes such as ballistic motion and diffusion. However, as we described 
in the previous sections, our formalism enables us to combine two states with 
arbitrary persistencies and describe the particle dynamics over all time scales. 
The general form of the expressions presented in the previous sections can be 
reduced to shorter formulas for specific choices of the two processes. Here we 
choose a diffusive dynamics ($\af{=}0$) for the dynamics of the slow state. 
The process can be further simplified by choosing constant velocities and 
also $\aFif\,{=}\,\af$ and $\aFfi\,{=}\,\ai$ at the switching events. Using 
these specific parameter choices leads to a reduced form of the MSD, as 
presented in Eq.\,(\ref{Eq:MSD-Diffusion_Persistent}). Then, the related 
transport quantities of interest can be extracted. The advantage of our 
formalism is that any desired feature of the motion can be kept in its 
general form. Particularly, the fast relocation mode is a persistent 
motion described by $\ai$ (and not a simple ballistic motion necessarily). 
For instance, we obtain from Eq.\,(\ref{Eq:Dasymp}) the following reduced 
form for the asymptotic diffusion coefficient in case of $\aFif\,{=}\,\af$ 
and $\aFfi\,{=}\,\ai$ and diffusive dynamics in state $\text{I\!I}$ ($\af{=}0$)
\begin{equation}
\begin{aligned}
D\!_{_\infty}{=}\frac{\Delta t}{4}\Bigg(&\frac{\ffi}{\fif{+}\ffi}\langle 
v^2\rangle_{_{\text{I}}}{+}\frac{\fif}{\fif{+}\ffi}\langle v^2\rangle_{_{
\text{I\!I}}} \\ 
&{-}2\frac{\fif\ffi\ai\langle v\rangle_{_{\text{I}}}\langle v\rangle_{_{
\text{I\!I}}}{+}(1{-}\fif)\ffi\ai\langle v\rangle_{_{\text{I}}}^2}{(\fif
{+}\ffi)\big(\ai(1{-}\fif){-}1\big)}\Bigg).
\end{aligned}
\label{Eq:DasympRT}
\end{equation}
For the explicit form of $D\!_{_\infty}$ in a ballistic-diffusive process, 
one readily replaces $\ai{=}1$ in the above equation. Note that 
Eq.\,(\ref{Eq:DasympRT}) also describes a combination of diffusion and 
subdiffusion for ${-1}{<}\ai{<}0$.    

{\it Optimization of transport quantities}--- We focused on the calculation 
of the displacement moments in this study. However, general conclusions 
may be also drawn for other transport quantities of interest, such as the 
mean-first-passage time (MFPT) to find a randomly located target. The MFPT 
is minimized in various biological systems to execute certain functions in an 
optimal way \cite{PerezIpina19,Bauer12,Schuss07,Bartumeus08}. Since the MFPT is 
conversely related to the asymptotic diffusion coefficient \cite{Condamin05,
Redner01}, achieving a minimum search time often corresponds to maximizing 
the diffusivity through $D\!_{_\infty}$. However, the optimization is only 
relevant with respect to those key factors that are accessible and can be 
varied by the biological agent. 

The advantage of having the explicit analytical form of the transport 
quantities of interest is that analytical expressions can be also extracted 
for the derivatives with respect to any control parameter, which makes the 
optimization of the transport quantities feasible. For example, the asymptotic 
diffusion coefficient given in Eq.\,(\ref{Eq:DasympRT}) can be optimized 
with respect to the switching probabilities using e.g.\ 
\begin{equation}
\frac{\partial D\!_{_\infty}}{\partial\fif}\Bigg|_{\fif=\fif^{\,\,
\text{opt}}}\!\!\!\!\!\!\!\!\!\!\!\!\!\!\!\!=0,
\label{Eq:Dderivative}
\end{equation}
to obtain the following relation between the optimal switching probabilities 
$\fif^{\,\,\text{opt}}$ and $\ffi^{\,\,\text{opt}}$ in a combination of ballistic 
($\ai{=}1$) and diffusive ($\af{=}0$) processes
\begin{equation}
\fif^{\,\,\text{opt}}\,{=}\,\displaystyle\frac{-2\,\ffi^{\,\,\text{opt}}\,
\langle v\rangle_{_{\text{I}}}^2+\ffi^{\,\,\text{opt}}\sqrt{B{-}2\,\langle 
v\rangle_{_{\text{I}}}^2}}{B},
\label{Eq:Doptimized}
\end{equation}
with $B{=}\displaystyle\ffi^{\,\,\text{opt}}\,(\langle v^2\rangle_{_{\text{I}}}
{-}2\langle v\rangle_{_{\text{I}}}^2){+}2\,\ffi^{\,\,\text{opt}}\,\langle 
v\rangle_{_{\text{I}}}\,\langle v\rangle_{_{\text{I\!I}}}{+}\langle 
v^2\rangle_{_{\text{I\!I}}}$. We find the necessary condition $\displaystyle 
B\,{\geq}\, 4\langle v\rangle_{_{\text{I}}}^4\,{+}\,2\langle 
v\rangle_{_{\text{I}}}^2$ for having an optimal solution. Figure~\ref{Fig:6}(b) 
shows a few optimal paths in the ($\fif,\ffi$) plane for various choices 
of velocity distribution in each state. 

\section{Conclusion}
\label{sec:conclusion}

We presented an analytical approach which provides a quantitative 
link between the characteristics of particle dynamics in a two-state 
active process and macroscopically observable transport properties. 
The method can be straightforwardly extended to three dimensions 
and multistate stochastic processes. We disentangled the combined 
effects of velocity, persistence, and switching statistics 
on the displacement moments. Importantly, the extracted explicit 
expressions for the MSD and related transport quantities such as the 
crossover time to long-term diffusion, initial anomalous exponent, and 
asymptotic diffusion coefficient (even for the simplified combination 
of a persistent walk and diffusion) reveal that the transport quantities 
of the multistate process cannot be simply obtained from the superposition 
of the individual states; in the presence of the products of the velocities 
or persistencies of the two states, or the product of the switching 
probabilities between the states, the transport quantities cannot 
be decomposed into pure contributions of the individual states. The 
extracted exact expression for the time evolution of the MSD and the 
detailed recipe to derive higher displacement moments make it possible 
to access first-passage and other transport quantities that can be 
expressed by a cumulant expansion in terms of the displacement moments. 
Alternatively, one may start with a master equation for the evolution 
of the quantity of interest similar to Eq.\,(\ref{Eq:MasterEqs}) and 
follow our analytical formalism to solve it. The presented approach 
is applicable to diverse transport problems in active matter systems 
as well as multistate passive processes such as clogging dynamics 
in granular media, chromatography, and transport in amorphous 
materials.

To be analytically tractable, we have considered noninteracting 
particles and only spontaneous transitions between the states (corresponding to 
exponential residence times in the states). Correlations and memory effects are 
not considered in the model presented in this study. However, the formalism is 
capable of handling correlations in general, e.g., by introducing aging for the 
switching probabilities (though one should then resort to numerical results for 
the transport quantities). While an analytical treatment of interacting persistent 
random walkers at the level of individual particles is unfeasible in general, 
the effects of the surrounding environment can be considered by effective 
turning-angle distributions via mean-field approaches. 

When the exerted forces on the active particle are known, the particle dynamics 
can be described by, e.g., Langevin or Fokker-Planck equations. However, if the 
exerted forces are unknown, our method proposes an alternative approach to obtain 
the macroscopically observable transport quantities of interest from the microscopic 
statistical properties of the particle trajectory. Our analytical formalism 
to describe the kinematics of active particles can describe stochastic processes 
in which external forces are replaced by their impact on the velocity and turning-angle 
distributions and the transition probabilities between the possible states. One can generalize 
this stochastic formalism and take other external fields, taxes, etc.\ into account 
through their influence on the movement and reorientation statistics of the particle. 

\section{Acknowledgments}
We acknowledge support from the Deutsche Forschungsgemeinschaft (DFG) 
through the collaborative research center SFB 1027. MRS acknowledges 
support by the Young Investigator Grant of the Saarland University, 
Grant No.\,7410110401.
\smallskip
\smallskip

\appendix

\section{Calculation of the displacement moments}
\label{Sec:Appendix}
In this appendix, we present the details of a Fourier-z-transform technique 
to extract analytical expressions for the displacement moments for the 
stochastic process described by the master equations (\ref{Eq:MasterEqs}). 
We adopted a matrix form in Eq.\,(\ref{Eq:MasterEqs}) to hint how the 
formalism can be generalized to multistate processes: One can consider 
$n$ states of motion and write the following set of master equations to 
link the states to each other
\begin{widetext}
\begin{align} \nonumber
&\SnewVec{=}\\ 
\,\,\,\,\,\,&\,\,\,\,\,\,\MtranNN\;\SoldVec\!. 
\label{Eq:MasterEqsGeneral}
\end{align}
Nevertheless, here we focus on the two-state process. The master equations 
(\ref{Eq:MasterEqs}) can be rewritten as  
\begin{eqnarray}
P_{\!_{t}}^{\text{I}}(x,y|\gamma) = \vspace{1mm} 
&&(1{-}f_{_{_{\text{I}\rightarrow\text{I\!I}}}}) 
\!\! \displaystyle\int \!\!\! dv F\!\!_{_{_{\text{I}}}}\!(v) \!\!
\displaystyle\int_{\!{-}\pi}^{\pi} \!\!\!\!\!\! d\beta \, R_{_{_{\text{I}}}}
\!(\gamma{-}\beta) P_{t{-}\Delta t}^{\text{I}}(x{-}v \Delta t\cos\gamma,y{-}
v \Delta t\sin\gamma|\beta) \vspace{1mm}\nonumber \\
&&+ f_{_{_{\text{I\!I}\rightarrow\text{I}}}} \!\! 
\displaystyle\int \!\!\! dv F\!\!_{_{_{\text{I}}}}\!(v) \!\!
\displaystyle\int_{\!{-}\pi}^{\pi} 
\!\!\!\!\!\! d\beta \, R_{_{_{\text{I\!I}\rightarrow\text{I}}}}\!(\gamma{-}\beta) 
\, P_{t{-}\Delta t}^{\text{I\!I}}(x{-}v \Delta t\cos\gamma,y{-}
v \Delta t\sin\gamma|\beta), \vspace{3mm}
\label{Eq:MasterEqs_Extended_Form_i}
\end{eqnarray}
\begin{eqnarray}
P_{\!_{t}}^{\text{I\!I}}(x,y|\gamma) = \vspace{1mm} 
&&(1{-}f_{_{_{\text{I\!I}\rightarrow\text{I}}}}) \!\!
\displaystyle\int \!\!\! dv F\!\!_{_{_{\text{I\!I}}}}\!(v) \!\!
\displaystyle\int_{\!{-}\pi}^{\pi} \!\!\!\!\!\! d\beta \, R_{_{_{\text{I\!I}}}}
\!(\gamma{-}\beta) P_{t{-}\Delta t}^{\text{I\!I}}(x{-}v \Delta t\cos\gamma,y{-}
v \Delta t\sin\gamma|\beta) \vspace{1mm}\nonumber \\
&&+ f_{_{_{\text{I}\rightarrow\text{I\!I}}}} \!\!
\displaystyle\int \!\!\! dv F\!\!_{_{_{\text{I\!I}}}}\!(v) \!\!
\displaystyle\int_{\!{-}\pi}^{\pi} 
\!\!\!\!\!\! d\beta \, R_{_{_{\text{I}\rightarrow\text{I\!I}}}}\!(\gamma{-}\beta) 
\, P_{t{-}\Delta t}^{\text{I}}(x{-}v \Delta t\cos\gamma,y{-}
v \Delta t\sin\gamma|\beta).
\label{Eq:MasterEqs_Extended_Form_ii}
\end{eqnarray}
\end{widetext}
It is unfeasible to solve the above set of equations in the general form 
to find the explicit form of the joint probability density function 
$P_{t}(x,y|\gamma)$. However, we prove in the following that exact analytical 
expressions can be obtained for arbitrary displacement moments. The 
Fourier transform of the probability density function in state $j$ is 
defined as
\begin{equation}
\mathcal{\widetilde{P}}_t^j({\bm k}|m)\,{=}\!\!\int\limits_{-\pi}^{\;\;\pi}
\!\!\text{d}\gamma\,e^{im\gamma} \!\!\int \!\! \text{d}y \!\!\int\!\!
\text{d}x\,\,e^{i\boldsymbol{k}\cdot\boldsymbol r} P_t^j(x,y|\gamma).
\label{Eq:P_Fourier_State_j}
\end{equation}
To obtain the Fourier transform of the master equations, we use the $g$-th 
order Bessel's function (with integer $g\,{\in}\,[{-}\infty,\infty]$)
\begin{equation}
J_g(z) = \frac{1}{2\pi i^{g}} \int_{-\pi}^{\pi} \!\! \text{d}\gamma \,\, 
e^{iz\cos\gamma} e^{-ig\gamma},
\end{equation}
and the Fourier transforms of the turning-angle distributions
\begin{eqnarray}
\nonumber
\mathcal{\widetilde{R}}_{_{\text{I}}}(m)&&\;{=}\!\displaystyle\int_{
-{\pi}}^{{\pi}}\!\!\!\!e^{i m \phi}\,R_{\text{I}}(\phi)\;\text{d}\phi, 
\\ \nonumber
\mathcal{\widetilde{R}}_{_{\text{I\!I}}}(m)&&\;{=}\!\displaystyle\int_{
-{\pi}}^{{\pi}}\!\!\!\!e^{i m \phi}\,R_{\text{I\!I}}(\phi)\;\text{d}\phi, 
\\ \nonumber
\mathcal{\widetilde{R}}_{_{\text{I}\rightarrow\text{I\!I}}}(m)&&\;
{=}\!\displaystyle\int_{-{\pi}}^{{\pi}}\!\!\!\!e^{i m \phi}\,R_{_{
\text{I}\rightarrow\text{I\!I}}}(\phi)\;\text{d}\phi, \\  
\mathcal{\widetilde{R}}_{_{\text{I\!I}\rightarrow\text{I}}}(m)&&\;
{=}\!\displaystyle\int_{-{\pi}}^{{\pi}}\!\!\!\!e^{i m \phi}\,R_{_{
\text{I\!I}\rightarrow\text{I}}}(\phi)\;\text{d}\phi.
\label{Eq:Fourier_Distributions}
\end{eqnarray}
Thus, the persistencies introduced in Eqs.\,(\ref{Eq:state_persistence}) 
and (\ref{Eq:switching_persistence}) are given as 
\begin{eqnarray}
\nonumber
\ai\,&&{=}\,\mathcal{\widetilde{R}}_{_{\text{I}}}(m\,{=}\,1), 
\\ \nonumber
\af\,&&{=}\,\mathcal{\widetilde{R}}_{_{\text{I\!I}}}(m\,{=}\,1), 
\\ \nonumber
\aFif\,&&{=}\,\mathcal{\widetilde{R}}_{_{\text{I}\rightarrow\text{I\!I}}}(m\,{=}\,1), 
\\  
\aFfi\,&&{=}\,\mathcal{\widetilde{R}}_{_{\text{I\!I}\rightarrow\text{I}}}(m\,{=}\,1).
\label{Eq:Fourier_Persistencies}
\end{eqnarray}
The master equations (\ref{Eq:MasterEqs_Extended_Form_i}) and 
(\ref{Eq:MasterEqs_Extended_Form_ii}) after Fourier transformation--- 
using the polar representation of $\boldsymbol{k}$ as $(k,\alpha)$--- 
read
\begin{equation}
\begin{aligned}
\mathcal{\widetilde{P}}_t^{\text{I}}(k, \alpha |m) = 
&\sum_{g=-\infty}^{\infty} \! i^g\,e^{-ig\alpha} \!
\displaystyle\int \!\!\! dv F\!\!_{_{_{\text{I}}}}\!(v) 
\,J_g(k \, v \Delta t) \times \\
&\Big[(1{-}f_{_{_{\text{I}{\!\rightarrow\!}\text{I\!I}}}})
\, \mathcal{\widetilde{R}}_{_{_{\text{I}}}}\!(m{+}g)\,\mathcal{
\widetilde{P}}_{t{-}\Delta t}^{\text{I}}(k,\alpha|m{+}g) \\
&+ f_{_{_{\text{I\!I}{\!\rightarrow\!}\text{I}}}} \, 
\mathcal{\widetilde{R}}_{_{_{\text{I\!I}\rightarrow\text{I}}}}\!(m{+}g) 
\, \mathcal{\widetilde{P}}_{t{-}\Delta t}^{\text{I\!I}}(k,\alpha |m{+}g)\Big],
\end{aligned}
\label{Eq:MasterEq_i_Fourier}
\end{equation}
\begin{equation}
\begin{aligned}
\mathcal{\widetilde{P}}_t^{\text{I\!I}}(k, \alpha |m) = 
&\sum_{g=-\infty}^{\infty} \! i^g\,e^{-ig\alpha} \!
\displaystyle\int \!\!\! dv F\!\!_{_{_{\text{I\!I}}}}\!(v) 
\,J_g(k \, v \Delta t) \times \\
&\Big[(1{-}f_{_{_{\text{I\!I}{\!\rightarrow\!}\text{I}}}})
\, \mathcal{\widetilde{R}}_{_{_{\text{I\!I}}}}\!(m{+}g)\,\mathcal{
\widetilde{P}}_{t{-}\Delta t}^{\text{I\!I}}(k,\alpha|m{+}g) \\
&+ f_{_{_{\text{I}{\!\rightarrow\!}\text{I\!I}}}} \, 
\mathcal{\widetilde{R}}_{_{_{\text{I}\rightarrow\text{I\!I}}}}\!(m{+}g) 
\, \mathcal{\widetilde{P}}_{t{-}\Delta t}^{\text{I}}(k,\alpha |m{+}g)\Big].
\end{aligned}
\label{Eq:MasterEq_ii_Fourier}
\end{equation}
The total probability density $\mathcal{\widetilde{P}}_t(k, \alpha |m)$ 
is then given by $\mathcal{\widetilde{P}}_t(k, \alpha |m)\,{=}\,
\mathcal{\widetilde{P}}_t^{\text{I}}(k, \alpha |m)\,{+}\,
\mathcal{\widetilde{P}}_t^{\text{I\!I}}(k, \alpha |m)$ and the 
displacement moments can be extracted as
\begin{equation}
\displaystyle\langle x^{a} y^{b} \rangle(t)= (-i)^{a+b} \frac{\partial^{a+b} 
\mathcal{\widetilde{P}}_t(k_{x},k_{y}|m=0)}{\partial k_{x}^{a}\partial k_{y}^{b}} 
\Big|_{(k_{x},k_{y})=(0,0)}.
\label{Eq:Moments_general}
\end{equation}
For example, the first four displacement moments along $x$ and 
$y$ directions are given by
\begin{align}
\nonumber
\langle x \rangle(t) &= \left. -i \frac{\partial 
\mathcal{\widetilde{P}}_t(k,\alpha{=}0|m{=}
0)}{\partial k}\right|_{k=0}, \\ \nonumber
\langle y \rangle(t) &= \left. -i \frac{\partial \mathcal{\widetilde{P}}_t\left(
k,\alpha{=}\frac{\pi}{2}\big|m{=}0\right)}{\partial k}
\right|_{k=0}, \\ \nonumber
\langle x^2 \rangle(t) &= \left. (-i)^2 \frac{\partial^2 
\mathcal{\widetilde{P}}_t(k,\alpha{=}0|m{=}
0)}{\partial k^2}\right|_{k=0}, \\ \nonumber
\langle y^2 \rangle(t) &= \left. (-i)^2 \frac{\partial^2 
\mathcal{\widetilde{P}}_t\left(k,\alpha{=}\frac{\pi}{2}\big|m{=}0\right)}{\partial 
k^2}\right|_{k=0}, \\ \nonumber
\langle x^3 \rangle(t) &= \left. (-i)^3 \frac{\partial^3 
\mathcal{\widetilde{P}}_t(k,\alpha{=}0|m{=}
0)}{\partial k^3}\right|_{k=0}, \\ \nonumber
\langle y^3 \rangle(t) &= \left. (-i)^3 \frac{\partial^3 
\mathcal{\widetilde{P}}_t\left(k,\alpha{=}\frac{\pi}{2}\big|m{=}0\right)}{\partial 
k^3}\right|_{k=0}, \\ \nonumber
\langle x^4 \rangle(t) &= \left. (-i)^4 \frac{\partial^4 
\mathcal{\widetilde{P}}_t(k,\alpha{=}0|m{=}0)}{\partial k^4}\right|_{k=0}, \\
\langle y^4 \rangle(t) &= \left. (-i)^4 \frac{\partial^4 
\mathcal{\widetilde{P}}_t\left(k,\alpha{=}\frac{\pi}{2}\big|m{=}0\right)}{\partial 
k^4}\right|_{k=0}.
\label{Eq:Moments_four_xy}
\end{align}
The Fourier transform of the probability in state $j$ can be expanded 
as a Taylor series
\begin{equation}
\begin{aligned}
\mathcal{\widetilde{P}}_{t}^j(k, \alpha|m) &= Q_{0,t}^j(\alpha|m) + i\,k
\!\! \displaystyle\int \!\!\! dv F\!\!_{_{_{\text{j}}}}(v) \, v \Delta t 
\,\, Q_{1,t}^j(\alpha|m) \\ 
&\hspace{4mm}- \frac{1}{2} k^2 
\!\! \displaystyle\int \!\!\! dv F\!\!_{_{_{\text{j}}}}(v) \, v^2 
(\Delta t)^2 \,\, Q_{2,t}^j(\alpha|m) \\
&\hspace{4mm}- \frac{i}{6} k^3 
\!\! \displaystyle\int \!\!\! dv F\!\!_{_{_{\text{j}}}}(v) \, v^3 
(\Delta t)^3 \,\, Q_{3,t}^j(\alpha|m) \\
&\hspace{4mm}+ \frac{1}{24} k^4 
\!\! \displaystyle\int \!\!\! dv F\!\!_{_{_{\text{j}}}}(v) \, v^4 
(\Delta t)^4 \,\, Q_{4,t}^j(\alpha|m) + \cdot \cdot \cdot,
\label{Eq:Taylor_Expansion}
\end{aligned}
\end{equation}
and the $h$-th displacement moment can be read in terms of the $h$-th 
Taylor expansion coefficient $Q_{h,t}^j(\alpha|m)$. The $x$ and $y$ 
components of the mean and the MSD in the state $j$ can be calculated as
\begin{align} 
\nonumber
\langle x \rangle^j(t) &=\! 
\displaystyle\int \!\!\! dv F\!\!_{_{_{\text{j}}}}(v) \,
v \Delta t \, Q_{1,t}^j(0|0),\\ \nonumber
\langle y \rangle^j(t) &= \!  
\displaystyle\int \!\!\! dv F\!\!_{_{_{\text{j}}}}(v) \,
v \Delta t \, 
Q_{1,t}^j\left(\frac{\pi}{2}\Big|0\right)\!,\\ \nonumber
\langle x^2 \rangle^j(t) &= \! 
\displaystyle\int \!\!\! dv F\!\!_{_{_{\text{j}}}}(v) \,
v^2 (\Delta t)^2 \, Q_{2,t}^j(0|0),\\
\langle y^2 \rangle^j(t) &= \!  
\displaystyle\int \!\!\! dv F\!\!_{_{_{\text{j}}}}(v) \,
v^2 (\Delta t)^2 \, Q_{2,t}^j\left(\frac{\pi}{2}\Big|0\right).
\label{Eq:First_Moments_Q}
\end{align}
One can similarly calculate higher displacement moments as well. 
For instance, the third and fourth moments are related to the 
Taylor expansion coefficients as 
\begin{align} 
\nonumber
\langle x^3 \rangle^j(t) &= \! 
\displaystyle\int \!\!\! dv F\!\!_{_{_{\text{j}}}}(v) \,
v^3 (\Delta t)^3 \, Q_{3,t}^j(0|0),\\ 
\nonumber
\langle y^3 \rangle^j(t) &= \!  
\displaystyle\int \!\!\! dv F\!\!_{_{_{\text{j}}}}(v) \,
v^3 (\Delta t)^3 \, Q_{3,t}^j\left(\frac{\pi}{2}\Big|0\right), \\ 
\nonumber
\langle x^4 \rangle^j(t) &= \! 
\displaystyle\int \!\!\! dv F\!\!_{_{_{\text{j}}}}(v) \,
v^4 (\Delta t)^4 \, Q_{4,t}^j(0|0),\\
\langle y^4 \rangle^j(t) &= \!  
\displaystyle\int \!\!\! dv F\!\!_{_{_{\text{j}}}}(v) \,
v^4 (\Delta t)^4 \, Q_{4,t}^j\left(\frac{\pi}{2}\Big|0\right).
\label{Eq:Higher_Moments_Q}
\end{align}
Thus, the problem reduces to the calculation of the Taylor 
expansion coefficients $Q_{h,t}^j(\alpha|m)$. We demonstrate 
in the following how $Q_{1,t}^j(\alpha|m)$ and $Q_{2,t}^j(\alpha|m)$ 
can be obtained, from which the mean and the MSD can be deduced. 
A similar procedure can be followed to extract higher expansion 
coefficients and, thus, higher displacement moments.

We expand both sides of the master equations (\ref{Eq:MasterEq_i_Fourier}) 
and (\ref{Eq:MasterEq_ii_Fourier}) and collect all terms with the 
same power in $k$. As a result, the following recursion relations 
for the Taylor expansion coefficients of terms with power $0$ in 
$k$ can be obtained
\begin{equation}
\begin{aligned}
&Q_{0,t}^{\text{I}}(\alpha|m)= \\
&(1{-}f_{_{_{\text{I}{\!\rightarrow\!}\text{I\!I}}}}) 
\mathcal{\widetilde{R}}_{_{_{\text{I}}}}\!(m) Q_{0,t{-}
\Delta t}^{\text{I}}(\alpha|m)+ f_{_{_{\text{I\!I}{\!\rightarrow\!}\text{I}}}} 
\mathcal{\widetilde{R}}_{_{_{\text{I\!I}{\!\rightarrow\!}\text{I}}}}\!(m) 
Q_{0,t{-}\Delta t}^{\text{I\!I}}(\alpha|m),
\end{aligned}
\label{Eq:Q_i_a}
\end{equation}
\begin{equation}
\begin{aligned}
&Q_{0,t}^{\text{I\!I}}(\alpha|m)= \\
&(1{-}f_{_{_{\text{I\!I}{\!\rightarrow\!}\text{I}}}}) 
\mathcal{\widetilde{R}}_{_{_{\text{I\!I}}}}\!(m) Q_{0,t{-}\Delta t}^{\text{I\!I}}
(\alpha|m)+ f_{_{_{\text{I}{\!\rightarrow\!}\text{I\!I}}}} 
\mathcal{\widetilde{R}}_{_{_{\text{I}{\!\rightarrow\!}\text{I\!I}}}}\!(m) 
Q_{0,t{-}\Delta t}^{\text{I}}(\alpha|m).
\end{aligned}
\label{Eq:Q_ii_a}
\end{equation}
Similarly, the expansion coefficients of terms with power $1$ in $k$ read
\begin{equation}
\begin{aligned}
Q_{1,t}^{\text{I}}(\alpha|m) =& (1{-}f_{_{_{\text{I}{\!\rightarrow\!}
\text{I\!I}}}})\bigg\{ \mathcal{\widetilde{R}}_{_{_{\text{I}}}}\!(m) 
Q_{1,t{-}\Delta t}^{\text{I}}(\alpha|m) \\
&{+}\frac{1}{2}\Big[ e^{i\alpha} \mathcal{\widetilde{R}}_{_{_{\text{I}}}}\!(m{-}1) 
Q_{0,t{-}\Delta t}^{\text{I}}(\alpha|m{{-}}1) \\
&{+} e^{{-}i\alpha} \mathcal{\widetilde{R}}_{_{_{\text{I}}}}\!(m{+}1)
Q_{0,t{-}\Delta t}^{\text{I}}(\alpha|m{{+}}1) \Big]\bigg\} \\
&{+} f_{_{_{\text{I\!I}{\!\rightarrow\!}\text{I}}}}
\bigg\{ \frac{\langle v \rangle_{_\text{I\!I}}}{\langle v \rangle_{_\text{I}}} 
\mathcal{\widetilde{R}}_{_{_{\text{I\!I}{\!\rightarrow\!}\text{I}}}}\!(m) 
Q_{1,t{-}\Delta t}^{\text{I\!I}}(\alpha|m) \\
&{+}\frac{1}{2}\Big[ e^{i\alpha} 
\mathcal{\widetilde{R}}_{_{_{\text{I\!I}{\!\rightarrow\!}\text{I}}}}\!(m{-}1) 
Q_{0,t{-}\Delta t}^{\text{I\!I}}(\alpha|m{{-}}1) \\ 
&{+} e^{{-}i\alpha} 
\mathcal{\widetilde{R}}_{_{_{\text{I\!I}{\!\rightarrow\!}\text{I}}}}\!(m{+}1) 
Q_{0,t{-}\Delta t}^{\text{I\!I}}(\alpha|m{{+}}1) \Big]\bigg\},
\end{aligned}
\label{Eq:Q_i_b}
\end{equation}
\begin{equation}
\begin{aligned}
Q_{1,t}^{\text{I\!I}}(\alpha|m) =& (1{-}f_{_{_{\text{I\!I}{\!\rightarrow\!}
\text{I}}}})\bigg\{ \mathcal{\widetilde{R}}_{_{_{\text{I\!I}}}}\!(m) 
Q_{1,t{-}\Delta t}^{\text{I\!I}}(\alpha|m) \\
&{+}\frac{1}{2}\Big[ e^{i\alpha} 
\mathcal{\widetilde{R}}_{_{_{\text{I\!I}}}}\!(m{-}1) 
Q_{0,t{-}\Delta t}^{\text{I\!I}}(\alpha|m{{-}}1) \\
&{+} e^{{-}i\alpha} \mathcal{\widetilde{R}}_{_{_{\text{I\!I}}}}\!(m{+}1)
Q_{0,t{-}\Delta t}^{\text{I\!I}}(\alpha|m{{+}}1) \Big]\bigg\} \\
&{+} f_{_{_{\text{I}{\!\rightarrow\!}\text{I\!I}}}}
\bigg\{ \frac{\langle v \rangle_{_\text{I}}}{\langle v \rangle_{_\text{I\!I}}} 
\mathcal{\widetilde{R}}_{_{_{\text{I}{\!\rightarrow\!}\text{I\!I}}}}\!(m) 
Q_{1,t{-}\Delta t}^{\text{I}}(\alpha|m) \\
&{+}\frac{1}{2}\Big[ e^{i\alpha} 
\mathcal{\widetilde{R}}_{_{_{\text{I}{\!\rightarrow\!}\text{I\!I}}}}\!(m{-}1) 
Q_{0,t{-}\Delta t}^{\text{I}}(\alpha|m{{-}}1) \\ 
&{+} e^{{-}i\alpha} 
\mathcal{\widetilde{R}}_{_{_{\text{I}{\!\rightarrow\!}\text{I\!I}}}}\!(m{+}1) 
Q_{0,t{-}\Delta t}^{\text{I}}(\alpha|m{{+}}1) \Big]\bigg\},
\end{aligned}
\label{Eq:Q_ii_b}
\end{equation}
and the expansion coefficients of terms with power $2$ in $k$ are
\begin{equation}
\begin{aligned}
Q_{2,t}^{\text{I}}(\alpha|m) &= \\
&\hspace{-15mm}(1{-}f_{_{_{\text{I}{\!\rightarrow\!}\text{I\!I}}}})
\bigg\{\Big[\frac{1}{2}Q_{0,t{-}\Delta t}^{\text{I}}(\alpha|m){+}
Q_{2,t{-}\Delta t}^{\text{I}}(\alpha|m)
\Big] \mathcal{\widetilde{R}}_{_{_{\text{I}}}}\!(m) \\
&{+} \frac{\langle v \rangle^2_{_\text{I}}}{\langle v^2 
\rangle_{_\text{I}}} \Big[e^{i
\alpha}Q_{1,t{-}\Delta t}^{\text{I}}(\alpha|m{-}1) 
\mathcal{\widetilde{R}}_{_{_{\text{I}}}}\!(m{-}1)\\
&{+}e^{{-}i\alpha}Q_{1,t{-}\Delta t}^{\text{I}}(\alpha|m{+}1) 
\mathcal{\widetilde{R}}_{_{_{\text{I}}}}\!(m{+}1)\Big]\\
&{+} \frac{1}{4}e^{2i\alpha}Q_{0,t{-}\Delta t}^{\text{I}}(\alpha|m{-}2) 
\mathcal{\widetilde{R}}_{_{_{\text{I}}}}\!(m{-}2)\\
&{+}\frac{1}{4}e^{{-}2i\alpha}Q_{0,t{-}\Delta t}^{\text{I}}(\alpha|m{+}2) 
\mathcal{\widetilde{R}}_{_{_{\text{I}}}}\!(m{+}2)\bigg\} \\
&\hspace{-15mm}+f_{_{_{\text{I\!I}{\!\rightarrow\!}\text{I}}}}\bigg\{
\Big[\frac{1}{2}Q_{0,t{-}\Delta t}^{\text{I\!I}}(\alpha|m){+} 
\frac{\langle v^2 \rangle_{_\text{I\!I}}}{\langle v^2 \rangle_{_\text{I}}} 
Q_{2,t{-}\Delta t}^{\text{I\!I}}(\alpha|m)\Big] 
\mathcal{\widetilde{R}}_{_{_{\text{I\!I}{\!\rightarrow\!}\text{I}}}}\!(m) \\
&{+} \frac{\langle v \rangle_{_\text{I}}\langle v 
\rangle_{_\text{I\!I}}}{\langle v^2 \rangle_{_\text{I}}} \Big[e^{i\alpha}
Q_{1,t{-}\Delta t}^{\text{I\!I}}(\alpha|m{-}1)
\mathcal{\widetilde{R}}_{_{_{\text{I\!I}{\!\rightarrow\!}\text{I}}}}\!(m{-}1)\\
&{+}e^{{-}i\alpha}Q_{1,t{-}\Delta t}^{\text{I\!I}}(\alpha|m{+}1)
\mathcal{\widetilde{R}}_{_{_{\text{I\!I}{\!\rightarrow\!}\text{I}}}}\!(m{+}1)\Big]\\
&{+} \frac{1}{4}e^{2i\alpha}Q_{0,t{-}\Delta t}^{\text{I\!I}}(\alpha|m{-}2)
\mathcal{\widetilde{R}}_{_{_{\text{I\!I}{\!\rightarrow\!}\text{I}}}}\!(m{-}2)\\
&{+}\frac{1}{4}e^{{-}2i\alpha}Q_{0,t{-}\Delta t}^{\text{I\!I}}(\alpha|m{+}2)
\mathcal{\widetilde{R}}_{_{_{\text{I\!I}{\!\rightarrow\!}\text{I}}}}\!(m{+}2)\bigg\}\!,
\end{aligned}
\label{Eq:Q_i_c}
\end{equation}
\begin{equation}
\begin{aligned}
Q_{2,t}^{\text{I\!I}}(\alpha|m) &= \\
&\hspace{-15mm}(1{-}f_{_{_{\text{I\!I}{\!\rightarrow\!}\text{I}}}})
\bigg\{\Big[\frac{1}{2}Q_{0,t{-}\Delta t}^{\text{I\!I}}(\alpha|m)
{+}Q_{2,t{-}\Delta t}^{\text{I\!I}}(\alpha|m)
\Big] \mathcal{\widetilde{R}}_{_{_{\text{I\!I}}}}\!(m) \\
&{+} \frac{\langle v \rangle^2_{_\text{I\!I}}}{\langle v^2 
\rangle_{_\text{I\!I}}} \Big[e^{i
\alpha}Q_{1,t{-}\Delta t}^{\text{I\!I}}(\alpha|m{-}1) 
\mathcal{\widetilde{R}}_{_{_{\text{I\!I}}}}\!(m{-}1)\\
&{+}e^{{-}i\alpha}Q_{1,t{-}\Delta t}^{\text{I\!I}}(\alpha|m{+}1) 
\mathcal{\widetilde{R}}_{_{_{\text{I\!I}}}}\!(m{+}1)\Big]\\
&{+} \frac{1}{4}e^{2i\alpha}Q_{0,t{-}\Delta t}^{\text{I\!I}}(\alpha|m{-}2) 
\mathcal{\widetilde{R}}_{_{_{\text{I\!I}}}}\!(m{-}2)\\
&{+}\frac{1}{4}e^{{-}2i\alpha}Q_{0,t{-}\Delta t}^{\text{I\!I}}(\alpha|m{+}2) 
\mathcal{\widetilde{R}}_{_{_{\text{I\!I}}}}\!(m{+}2)\bigg\} \\
&\hspace{-15mm}+f_{_{_{\text{I}{\!\rightarrow\!}\text{I\!I}}}}\bigg\{
\Big[\frac{1}{2}Q_{0,t{-}\Delta t}^{\text{I}}(\alpha|m){+} \frac{\langle 
v^2 \rangle_{_\text{I}}}{\langle v^2 \rangle_{_\text{I\!I}}} 
Q_{2,t{-}\Delta t}^{\text{I}}(\alpha|m)\Big] 
\mathcal{\widetilde{R}}_{_{_{\text{I}{\!\rightarrow\!}\text{I\!I}}}}\!(m) \\
&{+} \frac{\langle v \rangle_{_\text{I\!I}}\langle 
v \rangle_{_\text{I}}}{\langle v^2 \rangle_{_\text{I\!I}}} \Big[e^{i\alpha}
Q_{1,t{-}\Delta t}^{\text{I}}(\alpha|m{-}1)
\mathcal{\widetilde{R}}_{_{_{\text{I}{\!\rightarrow\!}\text{I\!I}}}}\!(m{-}1)\\
&{+}e^{{-}i\alpha}Q_{1,t{-}\Delta t}^{\text{I}}(\alpha|m{+}1)
\mathcal{\widetilde{R}}_{_{_{\text{I}{\!\rightarrow\!}\text{I\!I}}}}\!(m{+}1)\Big]\\
&{+} \frac{1}{4}e^{2i\alpha}Q_{0,t{-}\Delta t}^{\text{I}}(\alpha|m{-}2)
\mathcal{\widetilde{R}}_{_{_{\text{I}{\!\rightarrow\!}\text{I\!I}}}}\!(m{-}2)\\
&{+}\frac{1}{4}e^{{-}2i\alpha}Q_{0,t{-}\Delta t}^{\text{I}}(\alpha|m{+}2)
\mathcal{\widetilde{R}}_{_{_{\text{I}{\!\rightarrow\!}\text{I\!I}}}}\!(m{+}2)\bigg\}\!.
\end{aligned}
\label{Eq:Q_ii_c}
\end{equation}
Next, the time indices on both sides of the above equations can be 
equalized by means of $z$-transform, defined for the $h$-th Taylor 
expansion coefficient $Q_{h,t}^j(\alpha|m)$ as
\begin{equation}
\widehat{Q}_{h}^j(z,\alpha|m){=}\sum_{t{=}0}^\infty Q_{h,t}^j(\alpha|m) z^{-t}.
\label{Eq:z_transform_def}
\end{equation}
As a result, we obtain the expansion coefficients $\widehat{Q}_{h}^j(z,\alpha|m)$ 
of terms with power $h$ in the $z$-space. The $z$-transform of 
Eqs.\,(\ref{Eq:Q_i_a})-(\ref{Eq:Q_ii_c}) enables us to obtain 
the first two displacement moments in the $z$-space as
\begin{equation}
\begin{aligned}
\langle x \rangle(z) &{=} \sum_{t=0}^{\infty} z^{-t} \langle x 
\rangle(t) \\
&{=} \Delta t \Big[\langle v \rangle_{_\text{I}}\, 
\widehat{Q}_{1}^{\text{I}}(z,0|0) + \langle v \rangle_{_\text{I\!I}} 
\widehat{Q}_{1}^{\text{I\!I}}(z,0|0) \Big],
\end{aligned}
\end{equation}
and the second moment can be calculated as 
\begin{equation}
\begin{aligned}
\langle x^2 \rangle(z) &{=} \sum_{t=0}^{\infty} z^{-t} \langle x^2 
\rangle(t) \\
&{=} (\Delta t)^2 \Big[ \langle v^2 \rangle_{_\text{I}} 
\widehat{Q}_{2}^{\text{I}}(z,0|0) + \langle v^2 \rangle{_\text{I\!I}} 
\widehat{Q}_{2}^{\text{I\!I}}(z,0|0) \Big]. 
\end{aligned}
\end{equation}
For isotropic initial direction of motion the net displacement, i.e.\ 
the first moment, is zero. Thus, we carry the calculations in detail 
to extract the second displacement moment, i.e.\ the MSD, which 
is of particular interest. Using the $z$-transform of Eqs.\,(\ref{Eq:Q_i_c}) 
and (\ref{Eq:Q_ii_c}), $\langle x^2 \rangle(z)$ can be written in 
terms of $\widehat{Q}_{0}^j(z,0|0)$ and $\widehat{Q}_{1}^j(z,0|0)$ 
coefficients. Then, using Eq.\,(\ref{Eq:Fourier_Persistencies}) and 
the $z$-transform of Eqs.\,(\ref{Eq:Q_i_b}) and (\ref{Eq:Q_ii_b}), 
$\widehat{Q}_{1}^j(z,0|0)$ is eliminated to obtain
\begin{widetext}
\begin{equation}
\begin{aligned}
\langle x^2 \rangle(z) = &(\Delta t)^2 \Big[ 
(1{-}f_{_{_{\text{I}{\!\rightarrow\!}\text{I\!I}}}}) \widehat{Q}_{0}^{\text{I}}(z,0|0) 
{+} f_{_{_{\text{I\!I}{\!\rightarrow\!}\text{I}}}} \widehat{Q}_{0}^{\text{I\!I}}(z,0|0) 
\Big] \times \\
& \Bigg[ \frac{z\left[z{-}(1{-}f_{_{_{\text{I\!I}{\!\rightarrow\!}\text{I}}}})
a_{_{\text{I\!I}}}\right]}{(z{-}1)G(z)} \langle v \rangle^2_{_\text{I}} {+} 
\frac{z}{(z{-}1)G(z)} 
f_{_{_{\text{I}{\!\rightarrow\!}\text{I\!I}}}} \aFif 
\langle v \rangle_{_\text{I}} \langle v \rangle_{_\text{I\!I}} 
{-} \frac{1}{z{-}1} \langle v \rangle^2_{_\text{I}} {+} \frac{1}{2(z{-}1)} 
\langle v^2 \rangle_{_\text{I}} \Bigg] \\
&{+} (\Delta t)^2 \Big[f_{_{_{\text{I}{\!\rightarrow\!}\text{I\!I}}}} 
\widehat{Q}_{0}^{\text{I}}(z,0|0) 
{+} (1{-}f_{_{_{\text{I\!I}{\!\rightarrow\!}\text{I}}}}) 
\widehat{Q}_{0}^{\text{I\!I}}(z,0|0) \Big] \times \\
& \Bigg[ \frac{z\left[z{-}(1{-}f_{_{_{\text{I}{\!\rightarrow\!}\text{I\!I}}}})
a_{_{\text{I}}}\right]}{(z{-}1)G(z)} \langle v \rangle^2_{_\text{I\!I}} 
{+} \frac{z}{(z{-}1)G(z)} 
f_{_{_{\text{I\!I}{\!\rightarrow\!}\text{I}}}} \aFfi 
\langle v \rangle_{_\text{I\!I}} \langle v \rangle_{_\text{I}} 
{-} \frac{1}{z{-}1} \langle v \rangle^2_{_\text{I\!I}} {+} \frac{1}{2(z{-}1)} 
\langle v^2 \rangle_{_\text{I\!I}} \Bigg],
\end{aligned}
\end{equation}
\end{widetext}
by defining $G(z)=\left[z{-}(1{-}f_{_{_{\text{I\!I}{\!\rightarrow\!}\text{I}}}}) 
a_{_{\text{I\!I}}} \right]\left[z{-}(1{-}f_{_{_{\text{I}{\!\rightarrow\!}\text{I\!I}}}}) 
a_{_{\text{I}}} \right]-f_{_{_{\text{I}{\!\rightarrow\!}\text{I\!I}}}} 
f_{_{_{\text{I\!I}{\!\rightarrow\!}\text{I}}}} \aFif \aFfi$. Finally we replace 
$\widehat{Q}_{0}^{\text{I}}(z,0|0)$ and $\widehat{Q}_{0}^{\text{I\!I}}(z,0|0)$ from the $z$-transform 
of Eqs.\,(\ref{Eq:Q_i_a}) and (\ref{Eq:Q_ii_a}) to obtain $\langle x^2 \rangle(z)$ as
\begin{equation}
\begin{aligned}
&\langle x^2\rangle(z){=}\frac{(\Delta t)^2}{z{-}1}\!\displaystyle\sum_{\text{j}
{\in}\{\text{I},\text{I\!I}\},\text{j}{\neq}\text{j}'}\!\!\!\!\!\!\!\!\!\frac{z^2\fjpj\!\!
{+}(z^2{-}z)(1{-}\fjjp\!\!{-}\fjpj\!)q_{_0}^{\,\text{j}}}{\text{G}_{_0}(z)}
\!{\times}\\
&\Big[\!\frac{z\big[\!z{-}(1{-}\fjpj)\ajp\!\big]\!\vjm^2}{\text{G}_{_1}(z)} 
{+}\frac{z \fjjp\!a_{_{\text{j}\rightarrow\text{j}'}}\!\vjm\!\vjmp}{\text{G}_{_1}(z)}{-}\vjm^2
{+}\frac{\vjsm}{2}\!\Big],
\end{aligned}
\label{Eq:MSDz}
\end{equation}
where $\text{G}_{_1}(z)=\prod\limits_{\text{j}{\in}\{\text{I},\text{I\!I}\}}
\!\big[z{-}(1{-}\fjjp\!)\aj\big]{-}\prod\limits_{\text{j}{\in}\{\text{I},
\text{I\!I}\}}\!\fjjp\!a_{_{\text{j}\rightarrow\text{j}'}}$, 
$\text{G}_{_0}(z)\!\!=(z{-}1)(z{-}1{+}\ffi{+}\fif)$, and $q_{_0}^{\,\text{j}}$ 
is the probability of initially starting in state $\text{j}$. Note that $\Delta t$ 
can be absorbed into the velocity terms in the above equation to construct the 
mean step length $\langle \ell\rangle_{_{\text{j}}}\,{=}\,\langle v \Delta t
\rangle_{_{\text{j}}}$ or the second step-length moment $\langle \ell^2
\rangle_{_{\text{j}}}\,{=}\,\langle v^2(\Delta t)^2\rangle_{_{\text{j}}}$ 
in state $\text{j}$. By inverse $z$-transforming Eq.\,(\ref{Eq:MSDz}), an 
exact expression for $\langle x^2 \rangle(t)$ can be straightforwardly 
obtained. Since this expression is too lengthy, we define a couple of 
auxiliary quantities in the following to be able to present the explicit 
form of $\langle x^2 \rangle(t)$. For the isotropic initial direction of 
motion, we have $\langle x\rangle(t)\,{=}\,\langle y\rangle(t)\,{=}\,0$ 
and $\langle x^2\rangle(t)\,{=}\,\langle y^2\rangle(t)$; thus, $\langle 
r^2\rangle(t)$ can be obtained in 2D as $\langle r^2\rangle(t)\,{=}\,2
\langle x^2\rangle(t)$. By introducing
\begin{widetext}
\noindent$\Delta v_{_{\text{I}}}\,{=}\,\big(\Vdoi\,{-}\,2\,\vi^2\big)$, \\

\noindent$\Delta v_{_{\text{I\!I}}}\,{=}\,\big(\Vdof\,{-}\,2\,\vf^2\big)$, \\

\noindent$e_{_1}\,{=}\,(1\,{-}\,\fif)\,\ai$, \\

\noindent$e_{_2}\,{=}\,(1\,{-}\,\ffi)\,\af$, \\

\noindent$e_{_3}\,{=}\,\fif\,\ffi\,\aFif\,\aFfi$, \\

\noindent$e_{_4}\,{=}\,\fif\,\ffi\,(\aFif\,{+}\,\aFfi)$, \\

\noindent$e_{_5}\,{=}\,\sqrt{(e_{_1}{-}\,e_{_2})^2\,
{+}\,4\,e_{_3}}$\,, \\

\noindent$e_{_6}\,{=}\,(\fif{+}\ffi)\,\big(e_{_3}\,{-}\,
(e_{_1}{-}1)(e_{_2}{-}1)\big)$, 
\begin{eqnarray}
e_{_7}\,{=}\,&e_{_4}\Big((e_{_1}{+}e_{_2})(1{+}e_{_1}e_{_2}{-}e_{_3})
\,{+}\,4(e_{_3}{-}e_{_1}e_{_2})\,{+}\,e_{_5}(1{-}e_{_1}e_{_2}{+}e_{_3})\Big)
\vi\,\vf(\Delta t)^2 \hspace{98mm} \nonumber \\ 
&{+}\ffi\Big(e_{_1}(1{-}e_{_2})^2(-e_{_1}{+}e_{_2}{+}e_{_5})
+e_{_3}\big(3\,e_{_1}e_{_2}{+}2\,(e_{_2}{-}e_{_1})\big)-e_{_3}\big(e_{_2}^2
{+}e_{_2}e_{_5}{+}2\,(1{+}e_{_3}{+}e_{_5})\big)\Big)\vi^2 
(\Delta t)^2 \hspace{60.5mm} \nonumber \\
&{+}\fif\Big(e_{_2}(1{-}e_{_1})^2(-e_{_2}{+}e_{_1}{+}e_{_5})
+e_{_3}\big(3\,e_{_1}e_{_2}{+}2\,(e_{_1}{-}e_{_2})\big)-e_{_3}\big(e_{_1}^2
{+}e_{_1}e_{_5}{+}2\,(1{+}e_{_3}{+}e_{_5})\big)\Big)\vf^2 
(\Delta t)^2, \hspace{59.0mm} \nonumber
\end{eqnarray}
\begin{eqnarray}
e_{_8}\,{=}\,&e_{_4}\Big({-}(e_{_1}{+}e_{_2})(1{+}e_{_1}e_{_2}{-}e_{_3})
\,{-}\,4(e_{_3}{-}e_{_1}e_{_2})\,{+}\,e_{_5}(1{-}e_{_1}e_{_2}{+}e_{_3})\Big)
\vi\,\vf(\Delta t)^2 \hspace{98mm} \nonumber \\ 
&{+}\ffi\Big(e_{_2}(1{-}e_{_1})^2(-e_{_1}{+}e_{_2}{+}e_{_5})
-e_{_3}\big(3\,e_{_1}e_{_2}{+}2\,(e_{_2}{-}e_{_1})\big)-e_{_3}\big(e_{_1}^2
{-}e_{_1}e_{_5}{+}2\,(1{+}e_{_3}{+}e_{_5})\big)\Big)\vi^2 
(\Delta t)^2 \hspace{60.5mm} \nonumber \\
&{+}\fif\Big(e_{_1}(1{-}e_{_2})^2(-e_{_2}{+}e_{_1}{+}e_{_5})
-e_{_3}\big(3\,e_{_1}e_{_2}{+}2\,(e_{_1}{-}e_{_2})\big)-e_{_3}\big(e_{_2}^2
{-}e_{_2}e_{_5}{+}2\,(1{+}e_{_3}{+}e_{_5})\big)\Big)\vf^2 
(\Delta t)^2, \hspace{59.0mm} \nonumber
\end{eqnarray}
\begin{eqnarray}
e_{_9}\,{=}\,&\fif^{\,2}\,\ffi^{\,3}\,\aFif^2\,\aFfi^2\,\Delta\,v_{_{\text{I}}}
\,{+}\,\Big(\ffi(e_{_1}{-}1)^2(e_{_2}{-}1)^2\,{-}\,2\fif\ffi^{\,2}\aFif
\aFfi(e_{_1}{-}1)(e_{_2}{-}1)\Big)\,\Vdoi(\Delta t)^2 \hspace{98mm} \nonumber \\
&{+}\,\ffi^{\,2}\,\fif^{\,3}\,\aFif^2\,\aFfi^2\,\Delta\,v_{_{\text{I\!I}}} 
\,{+}\,\Big(\fif(e_{_1}{-}1)^2(e_{_2}{-}1)^2\,{-}\,2\ffi\fif^{\,2}\aFif
\aFfi(e_{_1}{-}1)(e_{_2}{-}1)\Big)\,\Vdof(\Delta t)^2 \hspace{95mm} \nonumber \\
&{+}\Big(2\fif\,\ffi^{\,2}\,\aFif\aFfi(3{-}2e_{_1}{-}2e_{_2}{+}2e_{_1}
e_{_2})\,{-}\,2\,\ffi\,e_{_1}(e_{_1}\,{-}\,2)(e_{_2}\,{-}\,1)^2\Big)\,\vi^2
(\Delta t)^2 \hspace{119mm} \nonumber \\
&{+}\Big(2\ffi\,\fif^{\,2}\,\aFif\aFfi(3{-}2e_{_1}{-}2e_{_2}{+}2e_{_1}
e_{_2})\,{-}\,2\,\fif\,e_{_2}(e_{_2}\,{-}\,2)(e_{_1}\,{-}\,1)^2\Big)\,\vf^2
(\Delta t)^2 \hspace{119mm} \nonumber \\
&{-}2\,\fif\,\ffi(\aFif{+}\aFfi)(e_{_1}{+}e_{_2}{-}2)\,\vi\,\vf(\Delta t)^2,
\hspace{165.8mm} \nonumber
\label{constant}
\end{eqnarray}

\noindent
$e_{_{10}}\,{=}\Big(\ffi(e_{_2}{-}1)\Vdoi{+}\fif(e_{_1}{-}1)\Vdof
{-}2\,e_{_4}\vi\vf{-}\ffi\big((e_{_2}{-}1)e_{_1}{-}e_{_3}\big)
\Delta\,v_{_{\text{I}}}{-}\fif\big((e_{_1}{-}1)e_{_2}{-}e_{_3}
\big)\Delta\,v_{_{\text{I\!I}}}\Big)(\Delta t)^2$, \\

\noindent we derive the following exact expression for the MSD
\begin{eqnarray}
\langle r^2\rangle(t)=\frac{e_{_9}+e_{_{10}}}{e_{_6}}+\frac{e_{_{10}}}{e_{_6}}\,t
+\frac{e_{_7}}{e_{_5}\,e_{_6}}\,\text{e}^{-t{/}\tcp}+\frac{e_{_8}}{e_{_5}\,e_{_6}}
\,\text{e}^{-t{/}\tcm},
\label{MSD_t}
\end{eqnarray}
with $t_{\text{c}_{\pm}}\,{=}\,{-}1{/}\ln(\frac{e_{_1}\,{+}\,e_{_2}\,
{\pm}\,e_{_5}}{2})$. For the combination of a persistent random walk 
and diffusion ($a_{_{\text{I\!I}}}{=}0$), constant velocities 
$v_{_{\text{I}}}\,{=}\,v_{_{\text{I\!I}}}\,{=}\,1$, and using 
$\aFif\,{=}\,\af$ and $\aFfi\,{=}\,\ai$ at the switching events, 
the MSD reduces to
\begin{equation}
\begin{aligned}
\langle r^2 \rangle(t) {=} &\frac{2\,a_{_{\text{I}}}\,
f_{_{_{\text{I}{\!\rightarrow\!}\text{I\!I}}}} (1{-}
f_{_{_{\text{I\!I}{\!\rightarrow\!}\text{I}}}}{-}
f_{_{_{\text{I}{\!\rightarrow\!}\text{I\!I}}}})^{t{+}
2}}{(f_{_{_{\text{I\!I}{\!\rightarrow\!}\text{I}}}}
{+}f_{_{_{\text{I}{\!\rightarrow\!}\text{I\!I}}}})^2 
(f_{_{_{\text{I\!I}{\!\rightarrow\!}\text{I}}}}{-}
f_{_{_{\text{I}{\!\rightarrow\!}\text{I\!I}}}} a_{_{\text{I}}}
{+}f_{_{_{\text{I}{\!\rightarrow\!}\text{I\!I}}}}{+}
a_{_{\text{I}}}{-}1)}{-} \frac{2\,a_{_{\text{I}}}\,
(f_{_{_{\text{I}{\!\rightarrow\!}\text{I\!I}}}}{-}1) 
\big(f_{_{_{\text{I}{\!\rightarrow\!}\text{I\!I}}}} 
a_{_{\text{I}}} (f_{_{_{\text{I\!I}{\!\rightarrow\!}\text{I}}}}
{+}f_{_{_{\text{I}{\!\rightarrow\!}\text{I\!I}}}}{-}2){+}
f_{_{_{\text{I\!I}{\!\rightarrow\!}\text{I}}}}{+}
f_{_{_{\text{I}{\!\rightarrow\!}\text{I\!I}}}}{+}
a_{_{\text{I}}}{-}1\big)\big(a_{_{\text{I}}}(1{-}
f_{_{_{\text{I}{\!\rightarrow\!}\text{I\!I}}}})\big)^t}{\big(
a_{_{\text{I}}}(f_{_{_{\text{I}{\!\rightarrow\!}\text{I\!I}}}}
{-}1){+}1\big)^2 (f_{_{_{\text{I\!I}{\!\rightarrow\!}\text{I}}}}
{-}f_{_{_{\text{I}{\!\rightarrow\!}\text{I\!I}}}} 
a_{_{\text{I}}}{+}f_{_{_{\text{I}{\!\rightarrow\!}\text{I\!I}}}}
{+}a_{_{\text{I}}}{-}1)} \\
&{+} \frac{a_{_{\text{I}}}^2 
(f_{_{_{\text{I}{\!\rightarrow\!}\text{I\!I}}}}{-}1) 
\Big((f_{_{_{\text{I\!I}{\!\rightarrow\!}\text{I}}}}{+}
f_{_{_{\text{I}{\!\rightarrow\!}\text{I\!I}}}}) 
\big(f_{_{_{\text{I\!I}{\!\rightarrow\!}\text{I}}}} 
(f_{_{_{\text{I}{\!\rightarrow\!}\text{I\!I}}}}{-}1)
{+}(f_{_{_{\text{I}{\!\rightarrow\!}\text{I\!I}}}}{-}3) 
f_{_{_{\text{I}{\!\rightarrow\!}\text{I\!I}}}}\big)
{+}2 f_{_{_{\text{I}{\!\rightarrow\!}\text{I\!I}}}}\Big)}{\big(
a_{_{\text{I}}}\,(f_{_{_{\text{I}{\!\rightarrow\!}\text{I\!I}}}}
{-}1)  (f_{_{_{\text{I\!I}{\!\rightarrow\!}\text{I}}}}{+}
f_{_{_{\text{I}{\!\rightarrow\!}\text{I\!I}}}})
{+}f_{_{_{\text{I\!I}{\!\rightarrow\!}\text{I}}}}{+}
f_{_{_{\text{I}{\!\rightarrow\!}\text{I\!I}}}}\big)^2} 
{-} \frac{2\,a_{_{\text{I}}}\big((
f_{_{_{\text{I\!I}{\!\rightarrow\!}\text{I}}}}{+}
f_{_{_{\text{I}{\!\rightarrow\!}\text{I\!I}}}})^2{-}
f_{_{_{\text{I}{\!\rightarrow\!}\text{I\!I}}}}\big)
{+}(f_{_{_{\text{I\!I}{\!\rightarrow\!}\text{I}}}}{+}
f_{_{_{\text{I}{\!\rightarrow\!}\text{I\!I}}}}
)^2}{\big(a_{_{\text{I}}}\,(
f_{_{_{\text{I}{\!\rightarrow\!}\text{I\!I}}}}{-}1) 
(f_{_{_{\text{I\!I}{\!\rightarrow\!}\text{I}}}}
{+}f_{_{_{\text{I}{\!\rightarrow\!}\text{I\!I}}}}){+}
f_{_{_{\text{I\!I}{\!\rightarrow\!}\text{I}}}}
{+}f_{_{_{\text{I}{\!\rightarrow\!}\text{I\!I}}}}\big)^2} \\
& {+} \frac{\Big(a_{_{\text{I}}} \big((
f_{_{_{\text{I\!I}{\!\rightarrow\!}\text{I}}}}{-}1) 
f_{_{_{\text{I}{\!\rightarrow\!}\text{I\!I}}}}{+}
f_{_{_{\text{I\!I}{\!\rightarrow\!}\text{I}}}}
{+}f_{_{_{\text{I}{\!\rightarrow\!}\text{I\!I}}}}^2\big)
{+}f_{_{_{\text{I\!I}{\!\rightarrow\!}\text{I}}}}
{+}f_{_{_{\text{I}{\!\rightarrow\!}\text{I\!I}}}}\Big)}{a_{_{\text{I}}}\,
(f_{_{_{\text{I}{\!\rightarrow\!}\text{I\!I}}}}
{-}1) (f_{_{_{\text{I\!I}{\!\rightarrow\!}\text{I}}}}
{+}f_{_{_{\text{I}{\!\rightarrow\!}\text{I\!I}}}})
{+}f_{_{_{\text{I\!I}{\!\rightarrow\!}\text{I}}}}
{+}f_{_{_{\text{I}{\!\rightarrow\!}\text{I\!I}}}}}
{+} \frac{\Big(a_{_{\text{I}}} 
\big((f_{_{_{\text{I\!I}{\!\rightarrow\!}\text{I}}}}{-}1) 
f_{_{_{\text{I}{\!\rightarrow\!}\text{I\!I}}}}
{+}f_{_{_{\text{I\!I}{\!\rightarrow\!}\text{I}}}}{+}
f_{_{_{\text{I}{\!\rightarrow\!}\text{I\!I}}}}^2\big)
{+}f_{_{_{\text{I\!I}{\!\rightarrow\!}\text{I}}}}{+}
f_{_{_{\text{I}{\!\rightarrow\!}\text{I\!I}}}}\Big)}
{a_{_{\text{I}}}\,(f_{_{_{\text{I}{\!\rightarrow\!}\text{I\!I}}}}{-}1) 
(f_{_{_{\text{I\!I}{\!\rightarrow\!}\text{I}}}}
{+}f_{_{_{\text{I}{\!\rightarrow\!}\text{I\!I}}}}){+}
f_{_{_{\text{I\!I}{\!\rightarrow\!}\text{I}}}}{+}
f_{_{_{\text{I}{\!\rightarrow\!}\text{I\!I}}}}} \,t.
\end{aligned}
\label{Eq:MSD-Diffusion_Persistent}
\end{equation}
\end{widetext}

\bibliography{Refs}

\end{document}